\documentclass[letterpaper,11pt]{article}

\usepackage{jheppub}

\usepackage{multirow}

\usepackage{subfig}
\usepackage{xspace}
\usepackage{xcolor}
\usepackage[countmax]{subfloat}
\usepackage{amsthm}

\newcommand{\theorem}{ansatz}

\DeclareRobustCommand{\Sec}[1]{Sec.~\ref{#1}}

\DeclareRobustCommand{\App}[1]{App.~\ref{#1}}

\DeclareRobustCommand{\Fig}[1]{Fig.~\ref{#1}}

\DeclareRobustCommand{\Eq}[1]{Eq.~(\ref{#1})}
\DeclareRobustCommand{\Eqs}[2]{Eqs.~(\ref{#1}) and (\ref{#2})}
\DeclareRobustCommand{\InRef}[1]{Ref.~\cite{#1}}
\DeclareRobustCommand{\Refs}[1]{Refs.~\cite{#1}}

\preprint{LA-UR-23-31203}

\title{
Flavor Fragmentation Function Factorization
}

\author[a]{Andrew J. Larkoski}
\author[b]{and Duff Neill}

\affiliation[a]{Department of Physics and Astronomy, University of California, Los Angeles, CA 90095, USA\\
Mani L. Bhaumik Institute for Theoretical Physics, University of California, Los Angeles, CA 90095, USA}
\affiliation[b]{Theoretical Division, Los Alamos National Laboratory, Los Alamos, New Mexico, 87545, USA}
\emailAdd{larkoski@ucla.edu}
\emailAdd{dneill@lanl.gov}

\abstract{
A definition of partonic jet flavor that is both theoretically well-defined and experimentally robust would have profound implications for measurements and predictions especially for heavy flavor applications.  Recently, a definition of jet flavor was introduced as the net flavor flowing along the direction of the Winner-Take-All axis of a jet which is soft safe to all orders, but not collinear safe.  Here, we exploit the lack of collinear safety and propose a factorization theorem of perturbative flavor fragmentation functions that resum collinear divergences and describe the evolution of flavor from the short distance of jet production to the long distance at which hadronization occurs.  Collinear flavor evolution is governed by a small modification of the DGLAP equations.  We present a detailed all-orders analysis and identify exact relations that must hold amongst the various anomalous dimensions by probability conservation and the existence of fixed points of the renormalization group flow.  We explicitly validate the factorization theorem at one-loop order, and demonstrate its consistency at two loops in particular flavor channels.  Starting at two-loops, constraints on phase space imposed by flavor measurements potentially allow for non-trivial soft contributions, but we demonstrate that they are scaleless and so explicitly vanish, ensuring that soft particles are summed inclusively and all divergences are exclusively collinear in nature.  This factorization theorem opens the door to precision calculations with identified flavor in the infrared.
}

\begin{document}
\maketitle

\section{Introduction}

Jets, collimated streams of high-energy particles and a foundational phenomena of quantum chromodynamics (QCD), provide a proxy for particle production at the shortest distances in particle collider experiments.  In addition to understanding the production of particle energy and momentum through jets, one would also desire an understanding of how other quantum numbers, such as electric charge or partonic flavor, are produced and then distributed throughout the numerous particles observed in experiment.  For electric charge, a practically useful definition of a jet's charge has been introduced long ago \cite{Field:1977fa}, and more recently has been put on a more firm theoretical foundation \cite{Waalewijn:2012sv,Chang:2013rca,Kang:2023ptt} enabling a richer quantitative interpretation.  Any observable sensitive to a jet's electric charge necessarily lacks the property of infrared and collinear (IRC) safety \cite{Sterman:1977wj,Ellis:1996mzs} that enables precision predictions exclusively within the perturbation theory of QCD.

Ascribing a partonic flavor to a jet is similar in many respects to electric charge, however it has been explicitly demonstrated that a jet's flavor can be defined in an entirely IRC safe way \cite{Banfi:2006hf}, as least on theoretical jets that consist exclusively of quarks and gluons, at a scale after parton showering but before hadronization.  Motivated by several recent high-precision calculations for the production of heavy-flavor jets \cite{Czakon:2020coa,Gauld:2020deh,Catani:2020kkl,Hartanto:2022ypo,Gauld:2023zlv}, many more jet flavor definitions have been proposed \cite{Caletti:2022glq,Caletti:2022hnc,Czakon:2022wam,Gauld:2022lem,Caola:2023wpj}.  For maximum utility within a high-precision fixed-order calculation, the jet flavor prescription must factorize from the kinematics of the jet and simply provide a label on an otherwise unflavored jet.  For applications to measurements at the Large Hadron Collider (LHC), this means that jet flavor should be a label to jets found with the anti-$k_T$ algorithm \cite{Cacciari:2008gp}.

With this as a goal, we follow the motivation of \InRef{Caletti:2022glq} and actually recommend that IRC safety {\it not} be a strict requirement for a useful jet flavor definition.  In particular, relaxing collinear safety but maintaining safety to exactly collinear emissions means that the jet flavor can be described by a flavor fragmentation function, whose evolution from high to low scales is completely governed by perturbatively-calculable differential equations.  \InRef{Caletti:2022glq} introduced this fragmentation function as defined by the net flavor that lies exactly on the Winner-Take-All recombination scheme axis of the jet \cite{gsalamwta,Bertolini:2013iqa,Larkoski:2014uqa}.  However, in that reference, the fragmentation function was introduced rather ad-hoc and resummation was only performed to leading-logarithmic (LL) accuracy, both of which are insufficient for applications to high-precision calculations.  In this paper, we will solve these issues and present a factorization \theorem~for jet flavor in which the flavor fragmentation function is but one part and well-defined to any perturbative order, as well as present first calculations necessary for next-to-leading logarithmic resummation and corresponding matching to next-to-next-to-leading fixed order.

This flavor factorization \theorem~will be presented within the context of factorization in soft-collinear effective theory (SCET) \cite{Bauer:2000yr,Bauer:2001ct,Bauer:2001yt,Bauer:2002nz}, with calculations performed in dimensional regularization and formulated as a product of functions each of which is defined at a unique scale.  Resummation is then accomplished by renormalization group evolution between the disparate scales of the functions in the factorization \theorem.  In a factorized form, the pieces that depend on the hard scattering process are separated from universal collinear functions, and so flavor fragmentation functions can be recycled from one calculation and used in numerous other contexts.  Within this framework, we reproduce the leading-logarithmic flavor evolution equations of \InRef{Caletti:2022glq} and extend their accuracy.  A central feature of this factorization \theorem~is that it is soft insensitive, as the Winner-Take-All recombination scheme axis is insensitive to soft particles. This we check explicitly via the zero-bin subtractions \cite{Manohar:2006nz}. The phase-space constraints are generally non-trivial when expanded in the zero-bin, but after appropriate rescalings, we can show the contributions are scaleless integrals.

Despite the significant evidence we present for this factorization \theorem, we emphasize again that we lack a more formal, rigorous proof, of identifying modes, their separation, and construction of a leading-power jet cross section as a product of matrix elements of SCET fields, as presented in, e.g., \Refs{Stewart:2010tn,Ellis:2010rwa,Bauer:2011uc,Becher:2015hka,Frye:2016aiz,Kang:2016mcy}.  At least as traditionally defined, collinear fields in SCET, those that are responsible for jet production, have a well-defined net partonic flavor and are invariant to collinear gauge transformations which requires that they can produce an arbitrary number of particles.  From the perspective of jet flavor advocated here, we would refer to the well-defined flavor of this SCET field as the ultraviolet (UV) flavor.  For measurements that are inclusive over emitted particles or insensitive to their flavor, this is preferred, because corresponding measurements that constrain the fields exclusively act on the distribution of momentum produced by the fields.  Requiring exclusive information about the flavor of particles produced in the jet would seem to modify the structure of and information carried by the fields.  We leave the question of feasibility and construction of IR flavor sensitive SCET fields to future work.

The outline of this paper is as follows.  In \Sec{sec:wtaflav}, we review the definition of jet flavor of \InRef{Caletti:2022glq} as the net flavor that flows exactly along the direction of the Winner-Take-All axis.  In \Sec{sec:factconj}, we present our conjecture for the factorization \theorem~of Winner-Take-All flavor, appropriate for jets produced in $e^+e^-$ collisions.  This is for simplicity here, to illustrate the structure and consistency of the factorization \theorem, but the extension to jets produced at a hadron collider is straightforward, as only a new hard function needs to be calculated.  In \Sec{sec:reneqanal} we present a general, all-orders analysis of the structure of UV renormalization from considerations like conservation of total flavor along the flow and the consequences of the symmetries of (perturbative) QCD.  This produces several non-trivial relationships between anomalous dimensions.  Exact, all-orders solutions to the renormalization group equations are presented in \Sec{sec:allordssols}, extending the leading-logarithmic predictions of \InRef{Caletti:2022glq} to whatever accuracy one is strong enough to calculate anomalous dimensions.  In \Sec{sec:uvrenormexp}, we present explicit calculations of the hard and jet flavor functions at one-loop order and demonstrate consistency of the renormalization structure implied by the factorization \theorem~at two-loops in \Sec{sec:as2calcs}.  In \Sec{sec:numerics}, we use these results to calculate the dependence of the Winner-Take-All flavor definition on the center-of-mass collision energy, resummed through LL$'$ accuracy. We conclude in \Sec{sec:concs} and discuss numerous applications of this factorization \theorem.  Appendices collect and summarize central results of the paper.

\section{Definition of WTA Jet Flavor}\label{sec:wtaflav}

In this section, we introduce the definition of the Winner-Take-All (WTA) flavor algorithm employed throughout the rest of this paper.  As our calculations will be restricted to jets produced in $e^+e^-$ collisions, we present the definition of WTA flavor for jets exclusively in the $e^+e^-$ collider environment.  Nevertheless, we note that the extension to WTA flavor appropriate for jets in a hadron collider is simply a change of coordinates in the reclustering metric and recombination scheme.  Because we only consider flavor measurements on already-clustered jets, there is no clustering with the beam, for example, that needs to be considered.  This further means that the algorithm of WTA flavor simply establishes a label for the jet given the flavors of its constituent partons, and in no way modifies any kinematic quantities of the jet.  As such, the kinematics of WTA flavored jets remains exactly that of the more familiar and experimentally-measured flavor-inclusive jets.

First, we define our jet sample in $e^+e^-$ collisions.  We will consider events divided into two hemisphere jets as defined by the plane transverse to the axis that minimizes thrust \cite{Brandt:1964sa,Farhi:1977sg}.  Then, on both of the hemisphere jets separately, we run the WTA flavor algorithm, as introduced in \InRef{Caletti:2022glq}:
\begin{enumerate}
\item Cluster and find jets in your collision event with any desired jet algorithm; in the case studied in this paper, defined through thrust hemispheres.
\item On a given jet, recluster its constituents with the $k_T$ algorithm \cite{Catani:1993hr,Ellis:1993tq}, using the WTA recombination scheme \cite{Bertolini:2013iqa,Larkoski:2014uqa,gsalamwta}.  Specifically:
\begin{enumerate}
\item For all pairs $i,j$ of particles in your jet, calculate the pairwise $k_T$ metric $d_{ij}=2\min[E_i^2,E_j^2](1-\cos\theta_{ij})$, where $E_i,E_j$ are the energies of particles $i,j$ and $\theta_{ij}$ is the angle between their three-momenta.
\item For the pair $i,j$ that corresponds to the smallest $d_{ij}$, recombine their momenta into a new massless particle $\widetilde{ij}$ such that $|\vec p_{\widetilde{ij}}| = |\vec p_i|+|\vec p_j|$, and the direction of $\widetilde{ij}$ is along the direction of the harder of $i$ and $j$.\footnote{For massless particles, this prescription is equivalent to setting the energy of particle $\widetilde{ij}$ to the sum of the daughter energies.  However, massive particles can be at rest which renders the energy form of the WTA recombination ambiguous, and so FastJet, for example, only implements the magnitude of three-momentum recombination rule \cite{Cacciari:2011ma}.  Throughout this paper, we only consider massless partons, so the magnitude of the three-momentum or energy form of the recombination are identical.}
\item Replace particles $i$ and $j$ with their combination $\widetilde{ij}$ in the collection of particles in the jet.
\item Repeat clustering until there is a single, combined particle that remains.  The direction of this particle corresponds to the direction of the WTA axis of the jet.
\end{enumerate}
\item The sum of the flavors of all particles in the jet whose momenta lie exactly along the WTA axis is defined to be the flavor of the jet.  This sum of flavors is defined as follows.  Gluons $g$ have 0 or neutral flavor.  A quark $q$ and anti-quark $\bar q$ of the same species have opposite flavor such that the flavor sum of a quark and its anti-quark is neutral (gluon): $q+\bar q = g$.  The flavor of two quarks of different species, say $q$ and $q'$, sum to a non-partonic flavor which we denote as $q+q'=(qq')$.
\end{enumerate}

In this algorithm, we follow the WTA axis all the way to the end of the parton shower, which is terminated at some physical scale $k_\perp^2 \sim 1 \text{ GeV}^2$, just before the scale of hadronization at which QCD becomes strongly coupled.  From a leading-order perspective, the WTA axis necessarily consists of a single parton, but at higher orders, in general, there will be the effect of unresolved emissions that lie within this cutoff scale, to ensure that the virtual and real divergences in the splitting functions that govern the parton shower cancel.  We will discuss how this scale arises and what emissions in the jets contribute at leading power to the WTA flavor in the following section.

As discussed in \InRef{Caletti:2022glq}, the WTA flavor algorithm is soft safe because soft particles cannot affect the location of the WTA axis.  However, WTA flavor is not collinear safe, because the WTA axis may move between different-flavored particles infinitesimally separated in angle, but is inclusive over {\it exactly} collinear splittings.  As we discuss more in \Sec{sec:hardfuncconsts}, inclusivity over exactly collinear splittings ensures that scale evolution of WTA flavor is governed by unambiguous parton-to-parton transitions, i.e., $u$ quark to $g$ gluon, which significantly simplifies the space of solutions. The lack of collinear safety means that the WTA flavor cannot be calculated in fixed-order perturbation theory, but strictly collinear unsafety means that divergences can be absorbed and resummed into appropriate fragmentation functions.  These fragmentation functions are defined through the factorization \theorem~that we present in the following section, but are much simpler than fragmentation functions that describe the transition of a perturbative parton to a non-perturbative hadron.  For flavor, the fragmentation function exclusively describes the evolution of a parton in the UV into the parton that lies along the WTA axis in the IR and so the WTA flavor fragmentation function can be completely calculated perturbatively.  That is, the anomalous dimensions of the evolution equations are perturbative as are the boundary conditions imposed on the flavor fragmentation function and so their solutions are completely perturbative.  However, they are not IRC safe, because there are interesting asymptotic limits in which a Taylor expansion about $\alpha_s=0$ breaks down \cite{Caletti:2022glq}, in a similar way to Sudakov safe observables studied in other contexts \cite{Larkoski:2013paa,Larkoski:2014wba,Larkoski:2014bia,Larkoski:2015lea}.

However, the perturbativity of the solutions to the evolution equations is a consequence of imposing a non-zero relative transverse momentum cut on the region about the WTA axis in which flavors are summed.  In the factorization conjecture that follows, we call this scale $k_\perp^2$, and further it is necessarily larger than the QCD scale, $k_\perp^2 \gg \Lambda_\text{QCD}^2$.  Within the context of a parton shower, $k_\perp^2$ would be comparable to the termination scale of the shower, before hadronization.  With these caveats, the analysis presented in this paper is the very first step toward systematic theoretical development of this flavor definition.  A practical and more realistic implementation of jet flavor will require inclusion of quark masses, as for heavy flavor applications, the mass of quarks is comparable to the scale of the end of the parton shower.  Further, partonic flavors are clearly not experimentally accessible, but augmentation of the factorization conjecture here with a non-perturbative function that describes evolution of partonic flavor from scale $k_\perp^2$ to hadronic flavor at $\Lambda_\text{QCD}^2$ may accomplish something toward this end.  We discuss these and other future directions more in the conclusions, \Sec{sec:concs}.

Additionally, throughout this paper, we exclusively consider the measurement of the WTA flavor label of jets, and no other flavor-sensitive information like the energy fraction distribution of the WTA axis.  A first study of kinematic quantities that are sensitive to an IR flavor definition was presented in \InRef{Caletti:2022glq}, and the analysis there could be extended.  However, we leave a more general analysis to future work, as the factorization of WTA flavor labels will be interesting and subtle enough.

\section{Factorization Conjecture}\label{sec:factconj}

Given the WTA flavor algorithm, we now present the conjecture for the factorization for its all-orders calculation in the collinear limit.  Here, we will apply the WTA flavor algorithm to both thrust hemisphere jets in $e^+e^-$ collisions, as described above.  We propose that the cross section $\sigma_{e^+e^-\to f_L f_R} $ for the scattered $e^+e^-$ to produce hemisphere WTA flavors $f_L f_R$ in the IR is
\begin{align}\label{eq:flavorfactthm}
\sigma_{e^+e^-\to f_L f_R} &=\sum_{j_L,\,j_R} H_{e^+e^-\to j_L j_R}(Q^2,\mu^2)\, J_{j_L\to f_L}(\mu^2,k_\perp^2)\, J_{j_R\to f_R}(\mu^2,k_\perp^2)\,.
\end{align}
In this expression, $Q^2$ is the squared collision energy or UV scale and $k_\perp^2$ is the scale of the termination of the parton shower according to the $k_T$ jet algorithm, at which the WTA flavor is measured.  This factorization \theorem~is valid in the limit in which the IR and UV scales are hierarchically separated, $Q^2/k_\perp^2 \to\infty$, with corrections expected as powers of $k_\perp^2/Q^2\ll 1$ in this limit.  $H_{e^+e^-\to j_L j_R}(Q^2,\mu^2)$ is the hard function that describes the production of hemisphere WTA flavor from the $e^+e^-$ collisions at scale $Q^2$ to the factorization scale $\mu^2$, at which the WTA flavors of the left and right hemispheres are $j_L, j_R$, respectively.  The jet function $J_{j_L\to f_L}(\mu^2,k_\perp^2)$ describes the evolution of the flavor of the left hemisphere from $j_L$ at scale $\mu^2$ down to the measurement scale $k_\perp^2$, at which the WTA flavor is $f_L$.  Note that the intermediate parton flavors $j_L,j_R$ that connect the hard function to the jet functions are summed over.

Evolution in $\mu^2$ between the hard function and either jet function proceeds via modified DGLAP evolution equations \cite{Gribov:1972ri,Gribov:1972rt,Lipatov:1974qm,Dokshitzer:1977sg,Altarelli:1977zs} that were first derived at leading-logarithmic accuracy in \Refs{Caletti:2022glq,Lifson:2020gua}.  The specific evolution equations will be presented in the following and anomalous dimensions evaluated a one-loop order, but their structure is to just follow the particle that lies on the WTA axis at every step in the evolution, as opposed to traditional DGLAP which follows the evolution of all particles produced at every step.  Additionally, we only track the flavor of the particle along the WTA axis, and no other moments of its energy, which renders the evolution equations especially simple, as coupled, linear differential equations.  We also note the probability-conservation sum rule, which, when all IR flavors are summed over, the total cross section that the $e^+e^-$ collision in the UV produced any pair of flavors is:
\begin{align}\label{eq:sumruletot}
\sum_{f_L,\,f_R}\sigma_{e^+e^-\to f_L f_R} = \sigma_\text{tot}\,,
\end{align}
where $\sigma_\text{tot}$ is the total hadronic cross section in $e^+e^-$ collisions.  We will exploit related sum rules for the hard and jet functions in the general analysis of the renormalization group equations in the following.

\subsection{Soft Effects}
In the infrared, we naively have contributions from both soft (s) and collinear (c) particles, whose momenta scale as:
\begin{align}
p_s&\sim (k_\perp,k_\perp,k_\perp)\,,\\
p_c &\sim\left(\frac{k_\perp^2}{Q},Q,k_\perp\right)\,.
\end{align}
The scale at which flavor is measured is a transverse momentum relative to the WTA axis of $k_\perp$, and so a soft particle is one for which its momentum scales like $k_\perp$ in all components.  That is, the soft momentum $p_s$ scales like $k_\perp$ in all lightcone coordinates relative to the WTA axis (the $z$-axis) where the first entry denotes the $+$-component ($p^+ = E-p_z$), the second entry is the $-$-component ($p^-=E+p_z$), and the final entry is the momentum transverse to the WTA axis. Such a momentum has energy that scales like $E_s\sim k_\perp$ and an angle relative to the WTA axis of $\theta_s \sim 1$. The collinear scaling implies the particles carry an order-1 fraction of the collision energy $Q$ and are confined to an angular region of size $\theta_c \sim k_\perp/Q\ll 1$.  These particles set the precise direction of the WTA axis, and are coherently recoiled by the soft emissions.

When practically calculating the jet functions in the factorization \theorem, we must impose the constraint on the relative transverse momentum of resolved particles.  This then defines the WTA flavor to be the net flavor in a region of size $k_\perp$, as defined by the $k_T$-algorithm. While soft particles cannot affect the WTA axis, they can be clustered within the scale $k_\perp$ of the WTA axis and therefore in principle affect the net flavor, according to this definition. However, we find that at the scale $k_\perp$, the soft contributions will be given by scaleless integrals, so that the collinear calculations will completely describe the UV renormalization of WTA flavor, through tracking the flavor of collinear particles along the renormalization group flow. That the soft contribution is scaleless is surprising as the naive zero-bin expansion \cite{Manohar:2006nz} of the relevant collinear phase-space appears to be non-trivial, but a more detailed analysis reveals the integrals to be scaleless.  This will be demonstrated explicitly at two-loop order in \Sec{sec:as2calcs}.

This does not imply that there is no soft contribution to the flavor of the jet. Strictly speaking, the net flavor of any region of finite size is ambiguous due to the presence of soft quarks beginning at order-$\alpha_s^2$ \cite{Banfi:2006hf}. However, the generation of soft flavor is therefore described by distinct evolution equations, exclusively at or below the scale $k_\perp$, which formally evolve the softs down to the non-perturbative QCD scale $\Lambda_{\rm QCD}$. Another way to describe the collinear flavor evolution that we study here is as the flavor that would be tracked to the end of a perturbative parton shower, from the collision scale $Q$ to the scale $k_\perp$, which represents the minimum scale that separates all resolved particles produced in the shower. Because no particles in a parton shower are produced within the relative scale $k_\perp$ of any other particle, the WTA flavor is unambiguous and has correspondingly no contribution from soft particles.  This understanding of WTA jet flavor could therefore be a relevant benchmark for high-order parton showers, such as, e.g., \Refs{Hamilton:2020rcu,Forshaw:2020wrq,Nagy:2020rmk,Herren:2022jej}, but we leave more details of such an analysis to future work.

\subsection{Ultraviolet Renormalization}\label{sec:UVrenorm}

As a first step of the analysis of this factorization \theorem, we present UV renormalization of the hard and jet functions and the definition of the renormalized objects which satisfy evolution equations.  For the hard function, because the WTA axis is insensitive to soft physics, there are no correlations between the particles that set the flavor in the two hemispheres, and so it is renormalized with a product of independent renormalization factors:
\begin{align}
H_{e^+e^-\to j_L j_R}^{\text{(bare)}}(Q^2,\mu_0^2) = \sum_{k_L,\, k_R}H_{e^+e^-\to k_L k_R}^\text{(ren)}(Q^2,\mu^2)\,Z_{k_L\to j_L}^{(H)}(\mu^2,\mu_0^2)\,Z_{k_R\to j_R}^{(H)}(\mu^2,\mu_0^2)\,.
\end{align}
The bare hard function is on the left which is independent of renormalization scale $\mu^2$, and the renormalized hard function is on the right, multiplied by renormalization factors.  Note the sum over intermediate flavors $k_L,\, k_R$ in the renormalization; renormalization of flavor is necessarily non-diagonal.  The jet functions are renormalized in a similar way, with, for the left hemisphere we have,
\begin{align}
J_{j_L\to f_L}^{\text{(bare)}}(\mu_0^2,k_\perp^2) = \sum_{k_L}Z_{j_L\to k_L}^{(J)}(\mu_0^2,\mu^2)\,J_{k_L\to f_L}^\text{(ren)}(\mu^2,k_\perp^2)\,.
\end{align}

Consistency of the factorization \theorem~requires independence of the flavor probabilities to the renormalization scale $\mu^2$.  Correspondingly, the product of the bare hard and jet functions is necessarily equal to the product of the renormalized hard and jet functions:
\begin{align}
&\sum_{j_L,\,j_R}H_{e^+e^-\to j_L j_R}^{\text{(bare)}}(Q^2,\mu_0^2)\, J_{j_L\to f_L}^{\text{(bare)}}(\mu_0^2,k_\perp^2)\, J_{j_R\to f_R}^{\text{(bare)}}(\mu_0^2,k_\perp^2) \\
&
\hspace{2cm}= \sum_{j_L,\, j_R}H_{e^+e^-\to j_L j_R}^\text{(ren)}(Q^2,\mu^2)\,J_{j_L\to f_L}^\text{(ren)}(\mu^2,k_\perp^2)\,J_{j_R\to f_R}^\text{(ren)}(\mu^2,k_\perp^2)\nonumber\,.
\end{align}
For this to hold, the hard and jet function renormalization factors are matrix inverses of one another:
\begin{align}
\sum_{j_L}Z_{i_L\to j_L}^{(H)}(\mu^2,\mu_0^2)\,Z_{j_L\to k_L}^{(J)}(\mu_0^2,\mu^2) = \delta_{i_L k_L}\,,
\end{align}
for the left hemisphere, and similar for the right hemisphere.  Then, for independence of the measurable flavor probabilities on the scale $\mu^2$, we must enforce the renormalization group evolution equations for the hard and jet functions:
\begin{align}\label{eq:uvrenorm}
&\mu^2\frac{\partial}{\partial \mu^2}H_{e^+e^-\to j_L j_R}^\text{(ren)}(Q^2,\mu^2) =- \sum_{k_L} H_{e^+e^-\to k_L j_R}^\text{(ren)}(Q^2,\mu^2)\,\gamma_{k_L\to j_L}-\sum_{k_R}H_{e^+e^-\to  j_L k_R}^\text{(ren)}(Q^2,\mu^2)\,\gamma_{k_R\to j_R}\,,\nonumber\\
&\mu^2\frac{\partial}{\partial \mu^2}J_{j_L\to f_L}^\text{(ren)}(\mu^2,k_\perp^2) =\sum_{k_L}\gamma_{j_L\to k_L}\, J_{k_L\to f_L}^\text{(ren)}(\mu^2,k_\perp^2)\,,\\
&\mu^2\frac{\partial}{\partial \mu^2}J_{j_R\to f_R}^\text{(ren)}(\mu^2,k_\perp^2) =\sum_{k_R}\gamma_{j_R\to k_R}\, J_{k_R\to f_R}^\text{(ren)}(\mu^2,k_\perp^2)\,.\nonumber
\end{align}
The $\gamma_{j\to k}$ factors are the UV anomalous dimensions and for which there are numerous all-orders constraints and relationships that we will review in the next section.  Also, while we have retained the superscript (ren) here to denote that these are the renormalized functions, we will typically not include the superscript working from the assumption that, unless otherwise noted, all functions are their renormalized versions.

\section{General Analysis of Renormalization Group Equations}\label{sec:reneqanal}

Here, we will provide several general, all-orders relationships between the functions in the factorization \theorem~and their anomalous dimensions.  The matrix structure of the hard and jet functions' evolution equations is particularly constraining.  We will study the hard and jet functions separately, and then provide a summary of the identities at the end of this section.

Some relationships between the anomalous dimensions that describe the evolution from WTA flavor $i$ to $j$ follow directly from symmetries of QCD itself.  By charge conjugation symmetry, the anomalous dimension from quark $q$ to anti-quark $\bar q$ is equal to the reverse process:
\begin{align}
\gamma_{q\to \bar q} = \gamma_{\bar q \to q}\,.
\end{align}
Charge conjugation also relates the transitions between gluons and (anti-)quarks:
\begin{align}
&\gamma_{g\to q} = \gamma_{g\to \bar q}\,, &\gamma_{q\to g} = \gamma_{\bar q\to g}\,.
\end{align}
Further, flavor symmetry between all massless species constrain anomalous dimensions involving a gluon to a quark, or vice-versa, to be independent of flavor:
\begin{align}
&\gamma_{g\to q} = \gamma_{g\to q'}\,, &\gamma_{q\to g} = \gamma_{q'\to g}\,,
\end{align}
for any two (anti-)quarks $q,q'$.  Finally, detailed balance requires that the evolution from a quark $q$ to another quark $q'$ is equivalent to the opposite process:
\begin{align}
\gamma_{q\to q'} = \gamma_{q'\to q}\,.
\end{align}
Flavor symmetry also requires that for $q'\neq q,\bar q$ that all other quark-to-quark transitions are described by a single anomalous dimension, $\gamma_{q\to q'}$.  Therefore, there are only six independent anomalous dimensions, regardless of the numbers of active quark flavors: $\gamma_{q\to q}$, $\gamma_{q\to \bar q}$,  $\gamma_{q\to q'}$, $\gamma_{g\to g}$, $\gamma_{g\to q}$, and $\gamma_{q\to g}$.

\subsection{Constraints from the Hard Function}\label{sec:hardfuncconsts}

Suppressing arguments, the hard function $H_{e^+e^-\to j_Lj_R}$ describes the evolution of WTA flavor from the center-of-mass collision scale $Q^2$ to the renormalization scale $\mu^2$.    At every step of this evolution, the flavor of the WTA axis necessarily changes into some flavor.  Further, this flavor is necessarily and unambiguously partonic, up quark, anti-strange quark, gluon, etc., because the WTA flavor in the hard function is determined by the net flavor in a region of 0 angular size about the WTA axis and exactly collinear splittings in QCD conserve partonic flavor.  Therefore, the hard function satisfies a sum rule, that the total probability that the initial scattered electron-positron pair evolves into any hemisphere WTA flavor pair $j_Lj_R$ is the total hadronic cross section:
\begin{align}
\sum_{j_L,j_R} H_{e^+e^-\to j_Lj_R}= \sigma_\text{tot}\,.
\end{align}

The renormalization group equations that govern the evolution of the hard function are, again,
\begin{align}
\mu^2\frac{\partial}{\partial \mu^2}H_{e^+e^-\to j_Lj_R} = - \sum_{k_L} H_{e^+e^-\to k_L j_R}\,\gamma_{k_L\to j_L}-\sum_{k_R}H_{e^+e^-\to  j_L k_R}\,\gamma_{k_R\to j_R}\,.
\end{align}
This is a set of linear, first-order differential equations and the anomalous dimensions $\gamma_{k\to j}$ are necessarily real numbers and independent of scale $\mu^2$ (except for implicit dependence in $\alpha_s$).  As such, the late-scale behavior in the limit that $\mu^2/Q^2\to 0$ of these equations is to necessarily flow to a fixed-point at which
\begin{align}
\lim_{\mu^2/Q^2\to 0}\mu^2\frac{\partial}{\partial \mu^2}H_{e^+e^-\to j_Lj_R} =0\,.
\end{align}
Because the equilibrium flavor is governed by the limit of collinear evolution, the equilibrium flavors of the two hemispheres are independent and the hard function reduces to a product of hemisphere hard functions, where
\begin{align}
H_{e^+e^-\to j_Lj_R} = \sigma_\text{tot}\,H_{j_L}H_{j_R}\,.
\end{align}
Therefore, in equilibrium, we need only study one hemisphere at a time.  With this property established, the relationship between the fixed-point fractions and the anomalous dimensions is simply
\begin{align}
\sum_{k}H_{k}\,\gamma_{k\to j}=0\,.
\end{align}

Relationships between the anomalous dimensions can be established by considering the specific final hemisphere flavor states, $j$.  If $j = q$, this equilibrium constraint is
\begin{align}
 0&=\sum_{k}H_{k}\,\gamma_{k\to q}=H_{q}\,\gamma_{q\to q}+H_{\bar q}\,\gamma_{\bar q\to q}+H_{g}\,\gamma_{g\to q}+\sum_{q'\neq q,\bar q} H_{q'}\,\gamma_{q'\to q}\,.
\end{align}
By the established symmetries of the anomalous dimensions, the equilibrium quark flavor fractions are independent of quark flavor, and so the hard functions involving only quarks are all equal:
\begin{align}
H_{q} = H_{\bar q}=H_{q'}\,.
\end{align}
Additionally, for $n_f$ active quark flavors, there are $2(n_f - 1)$ quarks $q'$ and so we find the relationship
\begin{align}\label{eq:equil1}
0=H_{q}\left[\gamma_{q\to q}+\gamma_{q\to \bar q}+2(n_f-1)\gamma_{q\to q'}\right]+H_{g}\,\gamma_{g\to q}\,.
\end{align}
Additionally, because of the independence of quark flavor on equilibrium fraction, the sum rule in equilibrium is simply
\begin{align}\label{eq:sumrule}
1 = \sum_q H_q+H_g = 2n_f H_q+H_g\,,
\end{align}
where on the right $H_q$ is the fraction that a single species of quark is the WTA flavor.

If instead the equilibrium flavor was a gluon, $j=g$, then the equilibrium constraint is
\begin{align}\label{eq:equil2}
0=\sum_{k}H_{k}\,\gamma_{k\to g}=\sum_qH_{q}\,\gamma_{q\to g}+H_{g}\,\gamma_{g\to g} = 2n_f H_q\gamma_{q\to g}+H_g \gamma_{g\to g}\,.
\end{align}
These three constraints, Eqs.~\ref{eq:equil1}, \ref{eq:sumrule}, and \ref{eq:equil2}, can be used to solve for the equilibrium flavor fractions where
\begin{align}
&H_q = \frac{-\frac{\gamma_{g\to g}}{2n_f}}{\gamma_{q\to g}-\gamma_{g\to g}}\,, &H_g = \frac{\gamma_{q\to g}}{\gamma_{q\to g}-\gamma_{g\to g}}\,.
\end{align}
This further produces the relationship between the anomalous dimensions where
\begin{align}\label{eq:hardanomconst}
2n_f\gamma_{g\to q}\gamma_{q\to g}=\gamma_{g\to g}\left[\gamma_{q\to q}+\gamma_{q\to \bar q}+2(n_f-1)\gamma_{q\to q'}\right]\,.
\end{align}

\subsection{Constraints from the Jet Function}

The most general form of the $\mu^2$ evolution equations for the jet function is of a similar form to that of the hard function, where
\begin{align}
\mu^2\frac{\partial}{\partial \mu^2} J_{j\to f} = \sum_k \gamma_{j\to k}J_{k\to f}\,,
\end{align}
again suppressing arguments for brevity.  Similar to the hard function, the jet function satisfies a probability-conservation sum rule, where, by summing over all possible WTA flavors,
\begin{align}
\sum_f J_{k\to f} = 1\,.
\end{align}
In the renormalization group evolution, summing over all WTA flavors then enforces that
\begin{align}
\mu^2\frac{\partial}{\partial \mu^2} \sum_fJ_{j\to f} =0= \sum_{k,f} \gamma_{j\to k}J_{k\to f} = \sum_k \gamma_{j\to k}\,,
\end{align}
and so the sum of the all possible anomalous dimensions for any UV flavor $j$ vanishes:
\begin{align}
\sum_k \gamma_{j\to k} = 0\,.
\end{align}
We can then consider the two cases of UV flavors, $q$ or $g$, and establish linear constraints on the anomalous dimensions.

Starting with a quark in the UV, the sum of anomalous dimensions is
\begin{align}\label{eq:quvjet}
0=\sum_k \gamma_{q\to k} = \gamma_{q\to q}+\gamma_{q\to \bar q}+2(n_f-1)\gamma_{q\to q'}+\gamma_{q\to g}\,,
\end{align}
where the quark flavor $q'\neq q,\bar q$.  For a gluon in the UV, the sum of anomalous dimensions is
\begin{align}\label{eq:guvjet}
0=\sum_k \gamma_{g\to k} = \gamma_{g\to g}+2n_f \gamma_{g\to q}\,.
\end{align}
We note that these two constraints, \Eqs{eq:quvjet}{eq:guvjet}, render the constraint from the hard function, \Eq{eq:hardanomconst}, a tautology.

\subsection{Summary of Relationships Between Anomalous Dimensions}\label{sec:anomsum}

We now summarize these general, all-orders relationships amongst the anomalous dimensions and fixed-point flavor fractions.  The only independent anomalous dimensions of UV renormalization are $\gamma_{q\to \bar q}$, $\gamma_{q\to q'}$, $\gamma_{q\to g}$, and $\gamma_{g\to q}$, for any quark flavor $q$ and $q'\neq q,\bar q$.  All other anomalous dimensions are related by the symmetries of QCD (flavor symmetry, charge conjugation, detailed balance) and the flavor-conserving anomalous dimensions are related by the existence of a fixed-point in the deep infrared or flavor-conservation:
\begin{align}
&\gamma_{q\to q} = -\gamma_{q\to \bar q}-2(n_f-1)\gamma_{q\to q'}-\gamma_{q\to g}\,, &\gamma_{g\to g} = -2n_f \gamma_{g\to q}\,,
\end{align}
for $n_f$ active quark flavors.  The simplest form of the infrared fixed-point flavors fractions for any quark species $q$ or gluon $g$ are
\begin{align}
&H_q  = \frac{\gamma_{g\to q}}{2n_f \gamma_{g\to q}+\gamma_{q\to g}}\,,
&H_g  = \frac{\gamma_{q\to g}}{2n_f \gamma_{g\to q}+\gamma_{q\to g}}\,.
\end{align}
These fixed points fractions are, in a general non-Abelian gauge theory, rational functions of the coupling, $\alpha_s$.  However, because of asymptotic freedom of QCD, in the limit that the physical scale $Q^2\to\infty$ for fixed renormalization scale $\mu^2$, these fixed-point fractions are independent of $\alpha_s$, where
\begin{align}
&H_q  =\lim_{\alpha_s\to 0} \frac{\gamma_{g\to q}(\alpha_s)}{2n_f \gamma_{g\to q}(\alpha_s)+\gamma_{q\to g}(\alpha_s)}\,,\\
&H_g  = \lim_{\alpha_s\to 0}\frac{\gamma_{q\to g}(\alpha_s)}{2n_f \gamma_{g\to q}(\alpha_s)+\gamma_{q\to g}(\alpha_s)}\,.
\end{align}
As the anomalous dimensions are a Taylor series in $\alpha_s$, the fixed point fractions therefore are actually exclusively determined by the one-loop anomalous dimensions.

\section{All-Orders Solution of Renormalization Group Evolution}\label{sec:allordssols}

Before presenting explicit calculations for the hard and jet functions at one-loop order, we will solve the general renormalization group equations to all orders.  This will illustrate their structure and the way in which the fixed points are approached.  For simplicity and compactness, we will only present the complete solution to the evolution equations for WTA gluon flavor, but more details and the solution for WTA quark flavor is presented in \App{app:quarkevowta}.

Again, the evolution equation for the jet function with identified WTA flavor $f$ is
\begin{align}
\mu^2\frac{\partial}{\partial \mu^2}J_{j\to f}(\mu^2,k_\perp^2) = \sum_k\gamma_{j\to k}J_{k\to f}(\mu^2,k_\perp^2) \,.
\end{align}
This can be re-expressed as evolution in the value of the coupling $\alpha_s$ through the $\beta$-function, where
\begin{align}
\mu^2\frac{\partial\alpha_s}{\partial\mu^2} = \beta(\alpha_s)\,.
\end{align}
Then, the renormalization group evolution is
\begin{align}
\frac{\partial}{\partial \alpha_s}J_{j\to f}\left(\alpha_s,\alpha_s(k_\perp^2)\right) = \frac{1}{\beta(\alpha_s)}\sum_k\gamma_{j\to k}(\alpha_s)\,J_{k\to f}\left((\alpha_s,\alpha_s(k_\perp^2)\right) \,,
\end{align}
with the boundary condition defined by the value of $\alpha_s$ at the scale $k_\perp^2$, $\alpha_s(k_\perp^2)$.

\subsection{Gluon WTA Flavor}

For gluon WTA flavor, this equation simplifies to a two-dimensional matrix equation
\begin{align}
\frac{\partial}{\partial \alpha_s}\left(\begin{array}{c}
J_{g\to g}\left(\alpha_s,\alpha_s(k_\perp^2)\right)\\
J_{q\to g}\left(\alpha_s,\alpha_s(k_\perp^2)\right)
\end{array}\right)=\frac{1}{\beta(\alpha_s)}\left(
\begin{array}{cc}
\gamma_{g\to g} &\sum_{q'}\gamma_{g\to q'}\\
\gamma_{q\to g} &\sum_{q'}\gamma_{q\to q'}
\end{array}
\right)\left(\begin{array}{c}
J_{g\to g}\left(\alpha_s,\alpha_s(k_\perp^2)\right)\\
J_{q\to g}\left(\alpha_s,\alpha_s(k_\perp^2)\right)
\end{array}\right)\,,
\end{align}
where $q'$ is any active quark flavor and we have used the relationships between anomalous dimensions.  Then, exploiting the fixed-point restrictions, this further simplifies to
\begin{align}
\frac{\partial}{\partial \alpha_s}\left(\begin{array}{c}
J_{g\to g}\left(\alpha_s,\alpha_s(k_\perp^2)\right)\\
J_{q\to g}\left(\alpha_s,\alpha_s(k_\perp^2)\right)
\end{array}\right)=\frac{1}{\beta(\alpha_s)}\left(
\begin{array}{cc}
\gamma_{g\to g} &-\gamma_{g\to g}\\
\gamma_{q\to g} &-\gamma_{q\to g}
\end{array}
\right)\left(\begin{array}{c}
J_{g\to g}\left(\alpha_s,\alpha_s(k_\perp^2)\right)\\
J_{q\to g}\left(\alpha_s,\alpha_s(k_\perp^2)\right)
\end{array}\right)\,,
\end{align}
The rate of evolution of the WTA flavor is determined by the eigenvalues of the anomalous dimension matrix.  Because of flavor conservation, this matrix has a 0 eigenvalue, while the other eigenvalue is
\begin{align}
\gamma_{g,\text{WTA}}\equiv \gamma_{g\to g}-\gamma_{q\to g}\,.
\end{align}
Therefore, this single eigenvalue controls the rate of evolution of the gluon WTA jet function from the scale $k_\perp^2$ to the scale $\mu^2$.

This matrix equation can be expanded out in terms of two coupled differential equations:
\begin{align}
\frac{\partial}{\partial\alpha_s}\left[
J_{g\to g}\left(\alpha_s,\alpha_s(k_\perp^2)\right)-J_{q\to g}\left(\alpha_s,\alpha_s(k_\perp^2)\right)
\right] &= \frac{\gamma_{g\to g}-\gamma_{q\to g}}{\beta(\alpha_s)}\left[
J_{g\to g}\left(\alpha_s,\alpha_s(k_\perp^2)\right)-J_{q\to g}\left(\alpha_s,\alpha_s(k_\perp^2)\right)
\right] \,,\nonumber\\
\frac{\partial}{\partial\alpha_s}\left[
J_{g\to g}\left(\alpha_s,\alpha_s(k_\perp^2)\right)+J_{q\to g}\left(\alpha_s,\alpha_s(k_\perp^2)\right)
\right] &= \frac{\gamma_{g\to g}+\gamma_{q\to g}}{\beta(\alpha_s)}\left[
J_{g\to g}\left(\alpha_s,\alpha_s(k_\perp^2)\right)-J_{q\to g}\left(\alpha_s,\alpha_s(k_\perp^2)\right)
\right] \,.
\end{align}
The first equation's solution is
\begin{align}
&J_{g\to g}\left(\alpha_s,\alpha_s(k_\perp^2)\right) =J_{g\to g}\left(\alpha_s(\mu_J^2),\alpha_s(k_\perp^2)\right)\, \exp\left[
\int_{\alpha_s(\mu_J^2)}^{\alpha_s}\frac{d\alpha}{\beta(\alpha)}\left(\gamma_{g\to g}(\alpha)-\gamma_{q\to g}(\alpha)\right)
\right]+F(\alpha_s)\,,\\
&J_{q\to g}\left(\alpha_s,\alpha_s(k_\perp^2)\right) =J_{q\to g}\left(\alpha_s(\mu_J^2),\alpha_s(k_\perp^2)\right)\, \exp\left[
\int_{\alpha_s(\mu_J^2)}^{\alpha_s}\frac{d\alpha}{\beta(\alpha)}\left(\gamma_{g\to g}(\alpha)-\gamma_{q\to g}(\alpha)\right)
\right]+F(\alpha_s)\,,
\end{align}
in terms of a natural scale $\mu_J^2$ of the jet function and for some function $F(\alpha_s)$.  The second differential equation in terms of $F(\alpha_s)$ can be written as
\begin{align}
\frac{\partial}{\partial \alpha_s}F(\alpha_s) &= \frac{\gamma_{q\to g}(\alpha_s)\,J_{g\to g}\left(\alpha_s(\mu_J^2),\alpha_s(k_\perp^2)\right)-\gamma_{g\to g}(\alpha_s)\,J_{q\to g}\left(\alpha_s(\mu_J^2),\alpha_s(k_\perp^2)\right)}{\beta(\alpha_s)}\\
&
\hspace{6cm}\times \exp\left[
\int_{\alpha_s(\mu_J^2)}^{\alpha_s}\frac{d\alpha}{\beta(\alpha)}\left(\gamma_{g\to g}(\alpha)-\gamma_{q\to g}(\alpha)\right)
\right]\nonumber\,.
\end{align}
Requiring that $F(\alpha_s)$ vanishes at the scale $\mu^2=\mu_J^2$, the solution can be written as
\begin{align}
F(\alpha_s) &=\int_{\alpha_s(\mu_J^2)}^{\alpha_s}d\alpha\,\frac{\gamma_{q\to g}(\alpha)\,J_{g\to g}\left(\alpha_s(\mu_J^2),\alpha_s(k_\perp^2)\right)-\gamma_{g\to g}(\alpha)\,J_{q\to g}\left(\alpha_s(\mu_J^2),\alpha_s(k_\perp^2)\right)}{\beta(\alpha)}\nonumber\\
&
\hspace{5cm}\times \exp\left[
\int_{\alpha_s(\mu_J^2)}^{\alpha}\frac{d\alpha'}{\beta(\alpha')}\left(\gamma_{g\to g}(\alpha')-\gamma_{q\to g}(\alpha')\right)
\right]\,.
\end{align}

Putting it together, the solutions for the jet functions are
\begin{align}\label{eq:gwtasolallords}
&J_{g\to g}\left(\alpha_s,\alpha_s(k_\perp^2)\right) =J_{g\to g}\left(\alpha_s(\mu_J^2),\alpha_s(k_\perp^2)\right)\, \exp\left[
\int_{\alpha_s(\mu_J^2)}^{\alpha_s}\frac{d\alpha}{\beta(\alpha)}\left(\gamma_{g\to g}(\alpha)-\gamma_{q\to g}(\alpha)\right)
\right]\nonumber\\
&\hspace{3cm}+\int_{\alpha_s(\mu_J^2)}^{\alpha_s}d\alpha\,\frac{\gamma_{q\to g}(\alpha)\,J_{g\to g}\left(\alpha_s(\mu_J^2),\alpha_s(k_\perp^2)\right)-\gamma_{g\to g}(\alpha)\,J_{q\to g}\left(\alpha_s(\mu_J^2),\alpha_s(k_\perp^2)\right)}{\beta(\alpha)}\nonumber\\
&
\hspace{5cm}\times \exp\left[
\int_{\alpha_s(\mu_J^2)}^{\alpha}\frac{d\alpha'}{\beta(\alpha')}\left(\gamma_{g\to g}(\alpha')-\gamma_{q\to g}(\alpha')\right)
\right]\,,\\
&J_{q\to g}\left(\alpha_s,\alpha_s(k_\perp^2)\right) =J_{q\to g}\left(\alpha_s(\mu_J^2),\alpha_s(k_\perp^2)\right)\, \exp\left[
\int_{\alpha_s(\mu_J^2)}^{\alpha_s}\frac{d\alpha}{\beta(\alpha)}\left(\gamma_{g\to g}(\alpha)-\gamma_{q\to g}(\alpha)\right)
\right]\nonumber\\
&\hspace{3cm}+\int_{\alpha_s(\mu_J^2)}^{\alpha_s}d\alpha\,\frac{\gamma_{q\to g}(\alpha)\,J_{g\to g}\left(\alpha_s(\mu_J^2),\alpha_s(k_\perp^2)\right)-\gamma_{g\to g}(\alpha)\,J_{q\to g}\left(\alpha_s(\mu_J^2),\alpha_s(k_\perp^2)\right)}{\beta(\alpha)}\nonumber\\
&
\hspace{5cm}\times \exp\left[
\int_{\alpha_s(\mu_J^2)}^{\alpha}\frac{d\alpha'}{\beta(\alpha')}\left(\gamma_{g\to g}(\alpha')-\gamma_{q\to g}(\alpha')\right)
\right]\,.
\end{align}
$\mu_J^2$ is the natural scale of the jet function and is taken to be of the order of $k_\perp^2$, but sensitivity to the specific choice of $\mu_J^2$ can be used to assess the residual scale dependence from a calculation truncated at a finite order.  We will show later that these expressions agree exactly with the results resummed at leading-logarithmic order in \InRef{Caletti:2022glq} with one-loop anomalous dimensions and $\beta$-function, and appropriate choices for the boundary conditions, $J_{g\to g}\left(\alpha_s(\mu_J^2),\alpha_s(k_\perp^2)\right)$ and $J_{q\to g}\left(\alpha_s(\mu_J^2),\alpha_s(k_\perp^2)\right)$.

\subsection{Quark WTA Flavor}\label{sec:qevoallords}

The evolution equations for WTA quark flavor are substantially more complicated, as described by a four-dimensional matrix equation:
\begin{align}
\frac{\partial}{\partial \alpha_s}\left(\begin{array}{c}
J_{q\to q}\left(\alpha_s,\alpha_s(k_\perp^2)\right)\\
J_{\bar q\to q}\left(\alpha_s,\alpha_s(k_\perp^2)\right)\\
J_{q'\to q}\left(\alpha_s,\alpha_s(k_\perp^2)\right)\\
J_{g\to q}\left(\alpha_s,\alpha_s(k_\perp^2)\right)
\end{array}\right)&=\frac{1}{\beta(\alpha_s)}\left(
\begin{array}{cccc}
\gamma_{q\to q} &\gamma_{q\to\bar q}&\sum_{q'}\gamma_{q\to q'} &\gamma_{q\to g}\\
\gamma_{\bar q\to q} &\gamma_{\bar q\to\bar q}&\sum_{q'}\gamma_{\bar q\to q'} &\gamma_{\bar q\to g}\\
\gamma_{q'\to q} &\gamma_{q'\to\bar q}&\sum_{q''}\gamma_{q'\to q''} &\gamma_{q'\to g}\\
\gamma_{g\to q} &\gamma_{g\to\bar q}&\sum_{q'}\gamma_{g\to q'} &\gamma_{g\to g}
\end{array}
\right)\left(\begin{array}{c}
J_{q\to q}\left(\alpha_s,\alpha_s(k_\perp^2)\right)\\
J_{\bar q\to q}\left(\alpha_s,\alpha_s(k_\perp^2)\right)\\
J_{q'\to q}\left(\alpha_s,\alpha_s(k_\perp^2)\right)\\
J_{g\to q}\left(\alpha_s,\alpha_s(k_\perp^2)\right)
\end{array}\right)\nonumber\\
&\hspace{-1cm}=\frac{1}{\beta(\alpha_s)}\left(
\begin{array}{cccc}
\gamma_{q\to q} &\gamma_{q\to\bar q}&2(n_f-1)\gamma_{q\to q'} &\gamma_{q\to g}\\
\gamma_{q\to \bar q} &\gamma_{q\to q}&2(n_f-1)\gamma_{q\to q'} &\gamma_{q\to g}\\
\gamma_{q\to q'} &\gamma_{q\to q'}&-\gamma_{q\to g}-2\gamma_{q\to q'} &\gamma_{q\to g}\\
\gamma_{g\to q} &\gamma_{g\to q}&2(n_f-1)\gamma_{g\to q} &-2n_f\gamma_{g\to q}
\end{array}
\right)\left(\begin{array}{c}
J_{q\to q}\left(\alpha_s,\alpha_s(k_\perp^2)\right)\\
J_{\bar q\to q}\left(\alpha_s,\alpha_s(k_\perp^2)\right)\\
J_{q'\to q}\left(\alpha_s,\alpha_s(k_\perp^2)\right)\\
J_{g\to q}\left(\alpha_s,\alpha_s(k_\perp^2)\right)
\end{array}\right)\,.
\end{align}
Here, the quark flavor $q'$ is distinct from $q$ and $\bar q$, $q' \neq q,\bar q$ and $q''$ is distinct from $q'$.  In writing these expressions, we have used the SU($n_f$) flavor symmetry and in the second line, have simplified the anomalous dimension matrix using the fixed-point relationships.  In this form, it is obvious that the anomalous dimension matrix has a 0 eigenvalue for total flavor conservation, while the three other eigenvalues are:
\begin{align}
\gamma_{q,\text{WTA}_1} &\equiv \gamma_{g\to g}-\gamma_{q\to g}\,,\\
\gamma_{q,\text{WTA}_2} &\equiv \gamma_{q\to q}-\gamma_{q\to\bar q}\,,\\
\gamma_{q,\text{WTA}_3} &\equiv -2n_f \gamma_{q\to q'}-\gamma_{q\to g}\,.
\end{align}
These three eigenvalues then control the rate of collinear evolution of the jet functions from the scale $k_\perp^2$ to the scale $\mu^2$.  For brevity here, we present the explicit solutions to these evolution equations with one-loop running in \App{app:quarkevowta}.

\section{Fixed-Order Analysis of UV Renormalization}\label{sec:uvrenormexp}

In this section, we present an explicit analysis of the UV evolution equations, evaluating the hard and jet functions at ${\cal O}(\alpha_s)$ to validate the consistency of the factorization \theorem, and presenting general expressions for the solution to the renormalization group equations through ${\cal O}(\alpha_s^2)$.  From these results, the established relationships between various flavor-changing anomalous dimensions can be verified and will set the stage for the calculation of new processes that start at ${\cal O}(\alpha_s^2)$ in \Sec{sec:as2calcs}.

\subsection{Solution to UV Renormalization Equations Through Order-$\alpha_s^2$}

We begin by presenting the general solution to the renormalization group equations through ${\cal O}(\alpha_s^2)$.  We will present the result explicitly for the jet function, but the expression for the hard function is similar.  We can expand the jet function in powers of $\alpha_s$ as
\begin{align}
J_{j\to f}(\mu^2,k_\perp^2) = \delta_{jf}+\frac{\alpha_s}{2\pi}\, f_{j\to f}^{(1)}\left(\log\frac{\mu^2}{k_\perp^2}\right)+\left(\frac{\alpha_s}{2\pi}\right)^2\, f_{j\to f}^{(2)}\left(\log\frac{\mu^2}{k_\perp^2}\right)+\cdots\,,
\end{align}
for some functions $f_{j\to f}^{(n)}$ of the logarithm of the ratio of scales at $n$th order in $\alpha_s$.  The running coupling satisfies the one-loop $\beta$-function
\begin{align}
\mu^2\frac{\partial \alpha_s}{\partial \mu^2} = -\frac{\alpha_s^2}{2\pi}\beta_0\,,
\end{align}
where $\beta_0$ is the one-loop coefficient,
\begin{align}
\beta_0 = \frac{11}{6}C_A - \frac{2}{3}T_Rn_f\,.
\end{align}
The anomalous dimension can also be expanded in powers of $\alpha_s$ as
\begin{align}
\gamma_{j\to k} = \frac{\alpha_s}{2\pi}\gamma_{j\to k}^{(1)}+\left(\frac{\alpha_s}{2\pi}\right)^2\gamma_{j\to k}^{(2)}+\cdots \,.
\end{align}

Then, the renormalization group evolution equation can be expressed through ${\cal O}(\alpha_s^2)$ as
\begin{align}
\mu^2\frac{\partial}{\partial \mu^2}J_{j\to f}(\mu^2,k_\perp^2) &=\frac{\alpha_s}{2\pi}\, \frac{d}{dL}f_{j\to f}^{(1)}(L)+\left(\frac{\alpha_s}{2\pi}\right)^2\,\left[ \frac{d}{dL}f_{j\to f}^{(2)}(L)-\beta_0\, f_{j\to f}^{(1)}(L)\right]\\
&=\sum_k\gamma_{j\to k}J_{k\to f}\nonumber\\
&=\frac{\alpha_s}{2\pi}\gamma^{(1)}_{j\to f}+\left(
\frac{\alpha_s}{2\pi}
\right)^2\left[
\sum_k\gamma^{(1)}_{j\to k}f^{(1)}_{k\to f}(L)+\gamma^{(2)}_{j\to f}
\right]\,,
\nonumber
\end{align}
where $L = \log\mu^2/k_\perp^2$.  Matching terms order-by-order, we find
\begin{align}
\frac{d}{dL}f_{j\to f}^{(1)}(L) = \gamma^{(1)}_{j\to f}\,,
\end{align}
or that
\begin{align}
f_{j\to f}^{(1)}(L) = \gamma^{(1)}_{j\to f}\, L+c^{(1)}_{j\to f}\,,
\end{align}
where $c^{(1)}_{j\to f}$ is a constant, independent of $\mu^2$.

With this result, the equation at the next order is
\begin{align}
\frac{d}{dL}f_{j\to f}^{(2)}(L)-\beta_0\, f_{j\to f}^{(1)}(L) = \sum_k\gamma^{(1)}_{j\to k}f^{(1)}_{k\to f}(L)+\gamma^{(2)}_{j\to f}\,,
\end{align}
or that
\begin{align}
\frac{d}{dL}f_{j\to f}^{(2)}(L)=\beta_0\,\gamma^{(1)}_{j\to f}\, L+\beta_0\,c_{j\to f}^{(1)}+\sum_k\left[\gamma^{(1)}_{j\to k}\, \gamma^{(1)}_{k\to f}\,L+\gamma^{(1)}_{j\to k}c_{k\to f}^{(1)}\right]+\gamma^{(2)}_{j\to f}\,.
\end{align}
The solution is
\begin{align}
f_{j\to f}^{(2)}(L) = \frac{\sum_k\gamma^{(1)}_{j\to k}\, \gamma^{(1)}_{k\to f}+\beta_0\,\gamma^{(1)}_{j\to f}}{2}\,L^2+\left(\beta_0c_{j\to f}^{(1)}+\sum_k\gamma^{(1)}_{j\to k}c_{k\to f}^{(1)}+\gamma^{(2)}_{j\to f}\right)L+c_{j\to f}^{(2)}\,,
\end{align}
where $c_{i\to f}^{(2)}$ is another constant.  Then, through ${\cal O}(\alpha_s^2)$, the jet function is
\begin{align}\label{eq:gensol2loop}
J_{j\to f}(\mu^2,k_\perp^2) &=\delta_{jf}+\frac{\alpha_s}{2\pi}\left( \gamma^{(1)}_{j\to f}\, \log\frac{\mu^2}{k_\perp^2}+c^{(1)}_{j\to f}\right)\\
&
\hspace{-2cm}+\left(
\frac{\alpha_s}{2\pi}
\right)^2\left(
  \frac{\sum_k\gamma^{(1)}_{j\to k}\, \gamma^{(1)}_{k\to f}+\beta_0\,\gamma^{(1)}_{j\to f}}{2}\,\log^2\frac{\mu^2}{k_\perp^2}+\left(\beta_0c_{j\to f}^{(1)}+\sum_k\gamma^{(1)}_{j\to k}c_{k\to f}^{(1)}+\gamma^{(2)}_{j\to f}\right)\log\frac{\mu^2}{k_\perp^2}+c_{j\to f}^{(2)}
 \right)\nonumber\\
 &\hspace{-2cm}+{\cal O}(\alpha_s^3)\,.\nonumber
\end{align}

\subsection{Order-$\alpha_s$ Calculations: Hard Function}

We now turn to explicit calculations of the jet and hard functions for WTA flavor in $e^+e^-$ collision events, starting with the hard function, $H_{e^+e^-\to j_Lj_R}$.  At leading-order, the hard function is diagonal, where
\begin{align}
H_{e^+e^- \to j_Lj_R} = \delta_{qj_L}\delta_{\bar q j_R}+\frac{\alpha_s}{2\pi}H^{(1)}_{e^+e^-\to j_L j_R}+{\cal O}(\alpha_s^2)\,,
\end{align}
where $q$ and $\bar q$ are an antiparticle pair of a quark $q$ and anti-quark $\bar q$.  We will calculate the one-loop corrections to this hard function, the factor $H^{(1)}_{e^+e^-\to j_L j_R}$, and through it will validate the general results established in previous sections.

First, the one-loop virtual contributions to the hard function exclusively contribute to the flavor-diagonal factors, where
\begin{align}
H_{e^+e^-\to q\bar q}^{(1,\text{V})} = C_F\left(
\frac{\mu^2}{Q^2}
\right)^\epsilon\left(
-\frac{2}{\epsilon^2}-\frac{3}{\epsilon}-8+\frac{7}{6}\pi^2
\right)\,,
\end{align}
which is the well-known one-loop virtual result for the $e^+e^-\to $ hadrons total cross section in dimensional regularization where the dimensionality of spacetime is $d = 4-2\epsilon$.

Real contributions to the hard function can result in flavor changes from the UV to the IR and can be calculated from the differential cross section for the $e^+e^- \to q\bar q g$ process, where
\begin{align}
\frac{1}{\sigma_0}\frac{d\sigma^{(1)}}{dx_1\, dx_2} &= \frac{\alpha_s }{2\pi}C_F\left(
\frac{\mu^2}{Q^2}
\right)^\epsilon\frac{1}{\Gamma(1-\epsilon)} \, (1-x_1)^{-\epsilon}\,(1-x_2)^{-\epsilon}\,(
x_1+x_2-1)^{-\epsilon}\\
&
\hspace{6cm}
\times
\frac{x_1^2+x_2^2-\epsilon(2-x_1-x_2)^2}{(1-x_1)(1-x_2)}
\,\Theta(x_1+x_2-1)\,.\nonumber
\end{align}
Here, $\sigma_0$ is the electroweak cross section for $e^+e^-\to q\bar q$ and the $x_1,x_2$ variables are the energy fractions of the quark and anti-quark with
\begin{align}
&x_1 = \frac{2E_q}{Q}\,, &x_2 = \frac{2E_{\bar q}}{Q}\,,
\end{align}
where $Q$ is the center-of-mass collision energy.  The energy fraction of the gluon is constrained by energy conservation, $x_3 = 2-x_1-x_2$.  The real emission contribution to the total cross section is well known, where
\begin{align}
\frac{\sigma^{(1,\text{R})}}{\sigma_0} = \frac{\alpha_s }{2\pi}C_F\left(
\frac{\mu^2}{Q^2}
\right)^\epsilon\left(
\frac{2}{\epsilon^2}+\frac{3}{\epsilon}+\frac{19}{2}-\frac{7}{6}\pi^2
\right)\,.
\end{align}
At this order, the only possible final hemisphere WTA flavor combinations are $q\bar q$, $qg$, and $g\bar q$ and so the sum of these three corresponding real contributions to the hard functions is constrained to equal this total:
\begin{align}
\frac{\alpha_s}{2\pi}\left(
H_{e^+e^-\to q\bar q}^{(1,\text{R})}+H_{e^+e^-\to qg}^{(1,\text{R})}+H_{e^+e^-\to g\bar q}^{(1,\text{R})}
\right)=\frac{\alpha_s }{2\pi}C_F\left(
\frac{\mu^2}{Q^2}
\right)^\epsilon\left(
\frac{2}{\epsilon^2}+\frac{3}{\epsilon}+\frac{19}{2}-\frac{7}{6}\pi^2
\right)\,.
\end{align}
Additionally, we note that, by charge conjugation, the hard functions $H_{e^+e^-\to qg}^{(1,\text{R})}=H_{e^+e^-\to g\bar q}^{(1,\text{R})}$, and so we only have to explicitly calculate the hard function $H_{e^+e^-\to q\bar q}^{(1,\text{R})}$.

The hemisphere WTA flavors are diagonal, $e^+e^-\to q\bar q$, if the gluon is the lowest energy particle, so that the WTA axes always lie on the quark and anti-quark.  The real contribution to the hard function is thus
\begin{align}
H_{e^+e^-\to q\bar q}^{(1,\text{R})}&=  C_F\left(
\frac{\mu^2}{Q^2}
\right)^\epsilon\frac{1}{\Gamma(1-\epsilon)}\int dx_1\, dx_2 \, (1-x_1)^{-\epsilon}\,(1-x_2)^{-\epsilon}\,(
x_1+x_2-1)^{-\epsilon}\\
&
\hspace{1cm}
\times
\frac{x_1^2+x_2^2-\epsilon(2-x_1-x_2)^2}{(1-x_1)(1-x_2)}
\,\Theta(x_1+x_2-1)\,\Theta(2x_1+x_2-2)\,\Theta(x_1+2x_2-2)\,
\nonumber\\
&=C_F\left(
\frac{\mu^2}{Q^2}
\right)^\epsilon\left(
\frac{2}{\epsilon^2}+\frac{3}{\epsilon}+\frac{-\frac{5}{4}+4\log 2}{\epsilon}+\frac{119}{24}-\frac{7}{6}\pi^2+\frac{7}{2}\log 2-\frac{5}{4}\log 3+4\log^2 2\right.\nonumber\\
&
\hspace{5cm}\left.+2\log^2 3+4\,\text{Li}_2\left(
\frac{2}{3}
\right)
\right)
\nonumber\,,
\end{align}
where $\text{Li}_2(x)$ is the dilogarithm function.  Summing this with the virtual contribution identified earlier, the one-loop flavor-diagonal hard function is
\begin{align}
H_{e^+e^-\to q\bar q}^{(1)} &=C_F\left(
\frac{\mu^2}{Q^2}
\right)^\epsilon\left(
\frac{-\frac{5}{4}+4\log 2}{\epsilon}-\frac{73}{24}+\frac{7}{2}\log 2-\frac{5}{4}\log 3+4\log^2 2+2\log^2 3+4\,\text{Li}_2\left(
\frac{2}{3}
\right)
\right)\,.
\end{align}
Importantly, all soft divergences have been explicitly canceled, and the remaining $1/\epsilon$ divergence is purely collinear and can therefore be removed by collinear renormalization described in \Sec{sec:UVrenorm}.  The hard functions $H_{e^+e^-\to qg}^{(1)}$ and $H_{e^+e^-\to g\bar q}^{(1)}$ are then
\begin{align}
H_{e^+e^-\to qg}^{(1)}&=H_{e^+e^-\to g\bar q}^{(1)}=\frac{\frac{2\pi}{\alpha_s}\frac{\sigma^{(1,\text{R})}}{\sigma_0}-H_{e^+e^-\to q\bar q}^{(1)}}{2}\\
&=C_F\left(
\frac{\mu^2}{Q^2}
\right)^\epsilon\left(
-\frac{-\frac{5}{8}+2\log 2}{\epsilon}+\frac{109}{48}-\frac{7}{4}\log 2+\frac{5}{8}\log 3-2\log^2 2-\log^2 3-2\,\text{Li}_2\left(
\frac{2}{3}
\right)
\right)
\nonumber\,.
\end{align}

From these results, we can then extract the one-loop anomalous dimensions $\gamma_{q\to q}^{(1)}$ and $\gamma_{q\to g}^{(1)}$.  Note that $\gamma_{q\to q}=\gamma_{\bar q\to \bar q}$, and so 
\begin{align}
\mu^2\frac{\partial }{\partial \mu^2}\frac{\alpha_s}{2\pi}H_{e^+e^-\to q\bar q}^{(1)} &= -\frac{\alpha_s}{2\pi}\sum_k H_{e^+e^-\to k\bar q}^{(0)}\gamma_{k\to q}^{(1)}-\frac{\alpha_s}{2\pi}\sum_k H_{e^+e^-\to qk}^{(0)}\gamma_{k\to \bar q}^{(1)}\\
&=-2\cdot\frac{\alpha_s }{2\pi}\,\gamma_{q\to q}^{(1)}\nonumber\\
& = 2\cdot  \frac{\alpha_s }{2\pi}C_F\left(
-\frac{5}{8}+2\log 2
\right)\,.\nonumber
\end{align}
The anomalous dimension $\gamma_{q\to g}^{(1)}$ is extracted from the evolution of the flavor changing process, for example
\begin{align}
\mu^2\frac{\partial}{\partial \mu^2}\frac{\alpha_s}{2\pi}H_{e^+e^-\to g\bar q}^{(1)} &= -\frac{\alpha_s}{2\pi}\sum_k H_{e^+e^-\to k\bar q}^{(0)}\gamma_{k\to g}-\frac{\alpha_s}{2\pi}\sum_k H_{e^+e^-\to gk}^{(0)}\gamma_{k\to \bar q} \\
&=-\frac{\alpha_s}{2\pi}\,\gamma_{q\to g}^{(1)}\nonumber\\
&= \frac{\alpha_s }{2\pi}C_F\left(
\frac{5}{8}-2\log 2
\right)\,.\nonumber
\end{align}
Note that this result establishes that the flavor-preserving and changing anomalous dimensions are opposite to one another at one-loop, $\gamma_{q\to g}^{(1)} = -\gamma_{q\to q}^{(1)}$.  This is consistent with the general relationship between these anomalous dimensions from \Sec{sec:anomsum} because all quark-flavor changing anomalous dimensions, $\gamma_{q\to q'}^{(1)}$, where $q'\neq q$, vanish at one-loop order.

\subsection{Order-$\alpha_s$ Calculations: Jet Function}

With the one-loop hard function calculated, we now turn to the one-loop jet function that describes collinear evolution of the flavor of the WTA axis.  We first review the procedure of renormalization of the jet function through one-loop, as these results will be useful for extracting two-loop anomalous dimensions and validating the general fixed-order expansion formula of the jet function in \Sec{sec:as2calcs}.

\subsubsection{Renormalization}\label{sec:jetrenorm}

From the renormalization prescription, the jet functions are explicitly renormalized as follows:
\begin{align}
J_{q\to q}^\text{(bare)} = Z_{q\to q} J_{q\to q}^\text{(ren)}+Z_{q\to g} J_{g\to q}^\text{(ren)}+\sum_{q''}Z_{q\to q''} J_{q''\to q}^\text{(ren)}\,,\\
J_{q\to g}^\text{(bare)} = Z_{q\to q} J_{q\to g}^\text{(ren)}+Z_{q\to g} J_{g\to g}^\text{(ren)}+\sum_{q''}Z_{q\to q''} J_{q''\to g}^\text{(ren)}\,,\\
J_{q\to q'}^\text{(bare)} = Z_{q\to q} J_{q\to q'}^\text{(ren)}+Z_{q\to g} J_{g\to q'}^\text{(ren)}+\sum_{q''}Z_{q\to q''} J_{q''\to q'}^\text{(ren)}\,\\
J_{g\to q}^\text{(bare)} = Z_{g\to q} J_{q\to q}^\text{(ren)}+Z_{g\to g} J_{g\to q}^\text{(ren)}+\sum_{q''}Z_{g\to q''} J_{q''\to q}^\text{(ren)}\,,\\
J_{g\to g}^\text{(bare)} = Z_{g\to q} J_{q\to g}^\text{(ren)}+Z_{g\to g} J_{g\to g}^\text{(ren)}+\sum_{q''}Z_{g\to q''} J_{q''\to g}^\text{(ren)}\,.
\end{align}
summed over intermediate quarks $q''\neq q$.  We have suppressed the arguments of these functions for brevity here as well.  At lowest order, $\alpha_s^0$, only the diagonal jet functions are non-zero, $J_{i\to j} = \delta_{ij}+{\cal O}(\alpha_s)$.  Therefore, the only non-zero renormalization factors are also diagonal:
\begin{align}
Z_{i\to j} = \delta_{ij}+{\cal O}(\alpha_s)\,.
\end{align}
Through ${\cal O}(\alpha_s)$, the quark flavor-changing jet functions are still 0: $J_{q''\to q} = {\cal O}(\alpha_s^2)$, and so through this order, the renormalized functions are
\begin{align}
J_{q\to q}^\text{(bare)} &= 1+Z^{(1)}_{q\to q}+ J_{q\to q}^\text{(1,ren)}+{\cal O}(\alpha_s^2)\,,\\
J_{q\to g}^\text{(bare)} &=  J_{q\to g}^\text{(1,ren)}+Z^{(1)}_{q\to g}+{\cal O}(\alpha_s^2)\,,\\
J_{q\to q'}^\text{(bare)} &= {\cal O}(\alpha_s^2)\,\\
J_{g\to q}^\text{(bare)} &= Z^{(1)}_{g\to q} +J_{g\to q}^\text{(1,ren)}+{\cal O}(\alpha_s^2)\,,\\
J_{g\to g}^\text{(bare)} &= 1+Z_{g\to g}^{(1)}+ J_{g\to g}^\text{(1,ren)}+{\cal O}(\alpha_s^2)\,.
\end{align}
Here, the superscript $1$ denotes the order in $\alpha_s$ at which these objects are calculated.

\subsubsection{$q\to q$ Jet Function Flavor Transition}

We now turn to the explicit calculation of the one-loop jet functions.  In this and all following subsections, we will denote the energy of the jet as $E_J$ and the relative transverse momentum between two particles produced in a splitting to be $\min[z^2,(1-z)^2]E_J^2\theta^2$, where $z,(1-z)$ are the energy fractions of the daughter particles and $\theta$ is their relative angle.  If the transverse momentum of the splitting is above some scale $k_\perp^2$, the flavor of the WTA axis is that of the more energetic particle.  If instead the transverse momentum of the splitting is below $k_\perp^2$, then the flavor of the WTA axis is the flavor of the initiating particle, which is also the sum of the flavors of the daughters.

With this set-up, the dimensionally-regulated, flavor-diagonal bare jet function $J^{(\text{bare})}_{q\to q}$ through one-loop is
\begin{align}
J^{\text{(bare)}}_{q\to q}(\mu_0^2,k_\perp^2) &=1+ \frac{\alpha_sC_F}{2\pi} \frac{(4\pi)^\epsilon}{\Gamma(1-\epsilon)}\left(
\frac{\mu_0^2}{E_J^2}
\right)^\epsilon\int_0^1dz\int_0^\infty d\theta^2 \\
&
\hspace{1cm}
\times (\theta^2)^{-1-\epsilon}z^{-2\epsilon}(1-z)^{-2\epsilon}\left(
\frac{1+(1-z)^2}{z}-\epsilon z
\right)\nonumber\\
&
\hspace{1cm}\times\left[\Theta(z^2\theta^2 E_J^2-k_\perp^2)\Theta(1/2-z)+\Theta(k_\perp^2-\min[z^2,(1-z)^2]\theta^2 E_J^2)\right]+{\cal O}(\alpha_s^2)\nonumber\\
&=1+\frac{\alpha_sC_F}{2\pi} \left(
\frac{\mu_0^2}{k_\perp^2}
\right)^\epsilon\frac{1}{\epsilon}\left[
\frac{5}{8}-2\log 2+\epsilon\left(
2-\frac{7}{4}\log 2-2\log^2 2
\right)
\right]+{\cal O}(\alpha_s^2)\,.\nonumber
\end{align}
To evaluate this expression, we have used the $q\to qg$ collinear splitting function and simplified the integrand at intermediate steps using the fact that scaleless integrals vanish in dimensional regularization.  Further, note that the zero-bin jet function to this order is scaleless,
\begin{align}
J^{\text{(zero,bare)}}_{q\to q}(\mu_0^2,k_\perp^2) &= \frac{\alpha_sC_F}{\pi} \frac{(4\pi)^\epsilon}{\Gamma(1-\epsilon)}\left(
\frac{\mu_0^2}{E_J^2}
\right)^\epsilon\int_0^\infty dz\int_0^\infty d\theta^2 \\
&
\hspace{1cm}
\times (\theta^2)^{-1-\epsilon}z^{-1-2\epsilon}(1-z)^{-2\epsilon}\nonumber\\
&
\hspace{1cm}\times\left[\Theta(z^2\theta^2 E_J^2-k_\perp^2)+\Theta(k_\perp^2-z^2\theta^2 E_J^2)\right]+{\cal O}(\alpha_s^2)\nonumber\\
&=0+{\cal O}(\alpha_s^2)\,,\nonumber
\end{align}
where we take the $z\sim k_\perp\to 0$ limit in the integrand.  

To eliminate divergences, we set the renormalization factor at this order to
\begin{align}
Z^{(1)}_{q\to q}(\mu_0^2,\mu^2) = \frac{\alpha_sC_F}{2\pi}\left(
\frac{\mu_0^2}{\mu^2}
\right)^\epsilon\frac{1}{\epsilon}\left(
\frac{5}{8}-2\log 2\right)\,,
\end{align}
so that the renormalized WTA jet function is
\begin{align}
J^\text{(ren)}_{q\to q}(\mu^2,k_\perp^2) &= J^\text{(bare)}_{q\to q}(\mu_0^2,k_\perp^2)-Z^{(1)}_{q\to q}(\mu_0^2,\mu^2)\\
&=1+\frac{\alpha_sC_F}{2\pi} \left(
\frac{\mu_0^2}{k_\perp^2}
\right)^\epsilon\frac{1}{\epsilon}\left[
\frac{5}{8}-2\log 2+\epsilon\left(
2-\frac{7}{4}\log 2-2\log^2 2
\right)
\right]\nonumber\\
&
\hspace{1cm}-\frac{\alpha_sC_F}{2\pi}\left(
\frac{\mu_0^2}{\mu^2}
\right)^\epsilon\frac{1}{\epsilon}\left(
\frac{5}{8}-2\log 2\right)+{\cal O}(\alpha_s^2)\nonumber\\
&=1+\frac{\alpha_s C_F}{2\pi}\left(
\left(
\frac{5}{8}-2\log 2
\right)\log\frac{\mu^2}{k_\perp^2}+2-\frac{7}{4}\log 2-2\log^2 2
\right)+{\cal O}(\alpha_s^2)\,.\nonumber
\end{align}
Its anomalous dimension is
\begin{align}
\gamma_{q\to q} = \frac{\alpha_s C_F}{2\pi}\left(
\frac{5}{8}-2\log 2
\right)+{\cal O}(\alpha_s^2)\,.
\end{align}
This is the same anomalous dimension as calculated from the hard function $H_{e^+e^-\to q\bar q}$.

\subsubsection{$q\to g$ Jet Function Flavor Transition}

The WTA flavor is only a gluon if there is a resolved emission and it is harder than the quark:
\begin{align}
J^{\text{(bare)}}_{q\to g}(\mu_0^2,k_\perp^2) &= \frac{\alpha_sC_F}{2\pi} \frac{(4\pi)^\epsilon}{\Gamma(1-\epsilon)}\left(
\frac{\mu_0^2}{E_J^2}
\right)^\epsilon\int_0^1dz\int_0^\infty d\theta^2 \\
&
\hspace{-1cm}
\times (\theta^2)^{-1-\epsilon}z^{-2\epsilon}(1-z)^{-2\epsilon}\left(
\frac{1+(1-z)^2}{z}-\epsilon z
\right)\,\Theta\left((1-z)^2\theta^2 E_J^2-k_\perp^2\right)\Theta(z-1/2)+{\cal O}(\alpha_s^2)\nonumber\\
&=\frac{\alpha_sC_F}{2\pi} \left(
\frac{\mu_0^2}{k_\perp^2}
\right)^\epsilon\frac{1}{\epsilon}\left[
-\frac{5}{8}+2\log 2+\epsilon\left(
-2+\frac{7}{4}\log 2+2\log^22
\right)
\right]+{\cal O}(\alpha_s^2)\,.\nonumber
\end{align}
For this jet function, the zero-bin explicitly vanishes because the gluon must be more energetic than the quark.  We set the renormalization factor to this order to
\begin{align}
Z_{q\to g}^{(1)}(\mu_0^2,\mu^2) = \frac{\alpha_sC_F}{2\pi}\left(
\frac{\mu_0^2}{\mu^2}
\right)^\epsilon\frac{1}{\epsilon}\left(
-\frac{5}{8}+2\log 2\right)\,,
\end{align}
so the renormalized WTA jet function is
\begin{align}
J^\text{(ren)}_{q\to g}(\mu^2,k_\perp^2) &= J^\text{(bare)}_{q\to g}(\mu_0^2,k_\perp^2)-Z^{(1)}_{q\to g}(\mu_0^2,\mu^2)\\
&=\frac{\alpha_sC_F}{2\pi} \left(
\frac{\mu_0^2}{k_\perp^2}
\right)^\epsilon\frac{1}{\epsilon}\left[
-\frac{5}{8}+2\log 2+\epsilon\left(
-2+\frac{7}{4}\log 2+2\log^22
\right)
\right]\nonumber\\
&
\hspace{1cm}- \frac{\alpha_sC_F}{2\pi}\left(
\frac{\mu_0^2}{\mu^2}
\right)^\epsilon\frac{1}{\epsilon}\left(
-\frac{5}{8}+2\log 2\right)+{\cal O}(\alpha_s^2)\nonumber\\ 
&=\frac{\alpha_s C_F}{2\pi}\left(
\left(
-\frac{5}{8}+2\log 2
\right)\log\frac{\mu^2}{k_\perp^2}-2+\frac{7}{4}\log 2+2\log^22
\right)+{\cal O}(\alpha_s^2)\,.\nonumber
\end{align}
Its anomalous dimension is then
\begin{align}
\gamma_{q\to g} = \frac{\alpha_s C_F}{2\pi}\left(
-\frac{5}{8}+2\log 2
\right)+{\cal O}(\alpha_s^2)\,.
\end{align}
This is the same anomalous dimension as calculated from the hard function $H_{e^+e^-\to g\bar q}$.

\subsubsection{$g\to g$ Jet Function Flavor Transition}

Now, onto an initial gluon.  For evolution into a gluon at the WTA flavor scale $k_\perp^2$, the contribution to the one-loop jet function from the $C_A$ collinear splitting channel is 0 because regardless of how the splitting occurs, there is perfect cancelation between real and virtual emissions.  However, there is a non-zero contribution from the $T_R$ channel if the splitting is unresolved:
\begin{align}
J^{(\text{bare})}_{g\to g}(\mu_0^2,k_\perp^2) &= 1+\frac{\alpha_s n_fT_R}{\pi} \frac{(4\pi)^\epsilon}{\Gamma(1-\epsilon)}\left(
\frac{\mu_0^2}{E_J^2}
\right)^\epsilon\int_0^1dz\int_0^\infty d\theta^2 \\
&
\hspace{0cm}
\times (\theta^2)^{-1-\epsilon}z^{-2\epsilon}(1-z)^{-2\epsilon}\left(
1-\frac{2z(1-z)}{1-\epsilon}
\right)\,\Theta(k_\perp^2-z^2\theta^2 E_J^2)\Theta(1/2-z)+{\cal O}(\alpha_s^2)\nonumber\\
&=1+\frac{\alpha_s n_fT_R}{2\pi} \frac{(4\pi)^\epsilon}{\Gamma(1-\epsilon)}\left(
\frac{\mu_0^2}{k_\perp^2}
\right)^\epsilon\frac{1}{\epsilon}\left[
-\frac{2}{3}+\epsilon\left(
-\frac{17}{18}+\frac{4}{3}\log 2
\right)
\right]+{\cal O}(\alpha_s^2)\,.\nonumber
\end{align}
The factor of $n_f$ appears because we must sum over all possible unresolved quarks.  We set the renormalization factor to this order to
\begin{align}
Z_{g\to g}^{(1)}(\mu_0^2,\mu^2) = -\frac{\alpha_s n_fT_R}{2\pi} \left(
\frac{\mu_0^2}{\mu^2}
\right)^\epsilon\frac{1}{\epsilon}
\frac{2}{3}\,,
\end{align}
so the renormalized WTA jet function is
\begin{align}
J^{(\text{ren})}_{g\to g}(\mu^2,k_\perp^2) &=J^\text{(bare)}_{g\to g}(\mu_0^2,k_\perp^2)-Z^{(1)}_{g\to g}(\mu_0^2,\mu^2)\\
&=1+\frac{\alpha_s n_fT_R}{2\pi} \left(
\frac{\mu_0^2}{k_\perp^2}
\right)^\epsilon\frac{1}{\epsilon}\left[
-\frac{2}{3}+\epsilon\left(
-\frac{17}{18}+\frac{4}{3}\log 2
\right)
\right] \nonumber\\
&\hspace{1cm}
+\frac{\alpha_s n_fT_R}{2\pi}\left(
\frac{\mu_0^2}{\mu^2}
\right)^\epsilon\frac{1}{\epsilon}
\frac{2}{3}+{\cal O}(\alpha_s^2)\nonumber\\
&=1+\frac{\alpha_s n_f T_R}{2\pi}\left(
-\frac{2}{3}\log\frac{\mu^2}{k_\perp^2}-\frac{17}{18}+\frac{4}{3}\log 2
\right)+{\cal O}(\alpha_s^2)\,.\nonumber
\end{align}
Its anomalous dimension is
\begin{align}
\gamma_{g\to g} =-\frac{2}{3} \frac{\alpha_s  n_fT_R}{2\pi}+{\cal O}(\alpha_s^2)\,.
\end{align}

\subsubsection{$g\to q$ Jet Function Flavor Transition}

For a quark from a gluon, this only occurs if the splitting is resolved and the quark takes away more energy than the anti-quark:
\begin{align}
J^{(\text{bare})}_{g\to q}(\mu_0^2,k_\perp^2) &= \frac{\alpha_sT_R}{2\pi} \frac{(4\pi)^\epsilon}{\Gamma(1-\epsilon)}\left(
\frac{\mu_0^2}{E_J^2}
\right)^\epsilon\int_0^{1/2}dz\int_0^\infty d\theta^2 \\
&
\hspace{1cm}
\times (\theta^2)^{-1-\epsilon}z^{-2\epsilon}(1-z)^{-2\epsilon}\left(
1-\frac{2z(1-z)}{1-\epsilon}
\right)\,\Theta(z^2\theta^2 E_J^2-k_\perp^2)+{\cal O}(\alpha_s^2)\nonumber\\
&=\frac{\alpha_sT_R}{2\pi} \frac{(4\pi)^\epsilon}{\Gamma(1-\epsilon)}\left(
\frac{\mu_0^2}{k_\perp^2}
\right)^\epsilon\frac{1}{\epsilon}\left[
\frac{1}{3}+\epsilon\left(
\frac{17}{36}-\frac{2}{3}\log 2
\right)
\right]+{\cal O}(\alpha_s^2)\,.\nonumber
\end{align}
We can then set the renormalization factor through this order to
\begin{align}
Z_{g\to q}^{(1)} = \frac{\alpha_sT_R}{2\pi} \left(
\frac{\mu_0^2}{\mu^2}
\right)^\epsilon\frac{1}{\epsilon}
\frac{1}{3}\,,
\end{align}
and the renormalized WTA jet function is
\begin{align}
J_{g\to q}(k_\perp) &= J^\text{(bare)}_{g\to q}(k_\perp)-Z^{(1)}_{g\to q}\\
&= \frac{\alpha_sT_R}{2\pi} \left(
\frac{\mu_0^2}{k_\perp^2}
\right)^\epsilon\frac{1}{\epsilon}\left[
\frac{1}{3}+\epsilon\left(
\frac{17}{36}-\frac{2}{3}\log 2
\right)
\right] -\frac{\alpha_sT_R}{2\pi} \left(
\frac{\mu_0^2}{\mu^2}
\right)^\epsilon\frac{1}{\epsilon}
\frac{1}{3}+{\cal O}(\alpha_s^2)\nonumber\\
&= \frac{\alpha_s T_R}{2\pi}\left(
\frac{1}{3}\log\frac{\mu^2}{k_\perp^2}+\frac{17}{36}-\frac{2}{3}\log 2
\right)+{\cal O}(\alpha_s^2)\,.\nonumber
\end{align}
Its anomalous dimension is
\begin{align}
\gamma_{g\to q} =\frac{1}{3} \frac{\alpha_s T_R}{2\pi}+{\cal O}(\alpha_s^2)\,.
\end{align}

\subsection{One-Loop Jet Function Summary}\label{sec:summoneloop}

We now summarize the results from calculating the one-loop jet functions for all possible flavor transistions.  The scaled anomalous dimensions are:
\begin{align}
&\gamma_{q\to q}^{(1)} = C_F\left(
\frac{5}{8}-2\log 2
\right)\,, &\gamma_{q\to g}^{(1)} = C_F\left(
-\frac{5}{8}+2\log 2
\right)\,,\\
&\gamma_{g\to q}^{(1)} = \frac{1}{3}T_R\,, &\gamma_{g\to g}^{(1)} = -\frac{2}{3}n_f T_R\,.\nonumber
\end{align}
These one-loop anomalous dimensions equal the corresponding results derived and used in \InRef{Caletti:2022glq} for one-loop evolution of the WTA flavor from the perspective of modifying DGLAP evolution.  The scaled one-loop constants of the jet function are
\begin{align}
&c_{q\to q}^{(1)} = C_F\left(
2-\frac{7}{4}\log 2-2\log^2 2
\right)\,, &c_{q\to g}^{(1)} = C_F\left(
-2+\frac{7}{4}\log 2+2\log^2 2
\right)\,,\\
&c_{g\to q}^{(1)} = T_R\left(
\frac{17}{36}-\frac{2}{3}\log 2
\right)\,, &c_{g\to g}^{(1)} = n_f T_R\left(
-\frac{17}{18}+\frac{4}{3}\log 2
\right)\,.\nonumber
\end{align}
These results are also collected in \App{app:oneloopsum}, along with the explicit expression for the hard function.

\section{Order-$\alpha_s^2$ Calculations with $q\to \bar q'q' q$ Matrix Element}\label{sec:as2calcs}

With results  at one-loop order established, we now move to studying the predictions and consequences of the WTA flavor factorization \theorem~at ${\cal O}(\alpha_s^2)$.  In this section, we will focus on flavor transitions in the jet function, validating that the factorization \theorem~and its renormalization procedure produces identical results to explicit calculation with complete matrix elements.  For all of these analyses, we focus on the collinear splitting of an initial quark $q$ into quarks of new flavors $q',\bar q'$, $q\to \bar q'_1q'_2 q_3$.  What is particularly interesting about this splitting is that there are multiple WTA flavor channels that are present.  This is the first order at which both quark flavor changes can occur, e.g., $q\to q'$, as well as non-partonic WTA flavor can be produced, e.g., $q\to (qq')$, where the two partons $q$ and $q'$ lie within a relative transverse momentum of $k_\perp^2$ of each other.  Crucially, we will demonstrate that the zero-bin is indeed scaleless and the resulting jet functions have no IR divergences, which also ensures that the non-partonic WTA flavor jet function has no UV divergences at this order.

The bare jet function for transition from the initial quark $q$ to a final flavor $f$ can be calculated from this splitting via the master formula
\begin{align}\label{eq:twoloopjetmaster}
J^{(2,\text{bare})}_{q\to f}(\mu_0^2,k_\perp^2) = \int d\Phi_3\, |{\cal M}_{q\to \bar q'_1q'_2 q_3}|^2\,\Theta_\text{WTA}\,.
\end{align}
$d\Phi_3$ is differential three-body collinear phase space in $d=4-2\epsilon$ dimensions.  Explicitly, this is \cite{Gehrmann-DeRidder:1997fom,Ritzmann:2014mka}
\begin{align}\label{eq:3bodps}
d\Phi_3 = \frac{4^{1-\epsilon}E_J^4}{(4\pi)^{5-2\epsilon}}\frac{(E_J^2)^{-2\epsilon}}{\Gamma(1-2\epsilon)}\,d\theta_{12}^2\,d\theta^2_{13}\, d\phi\,dz_1\,dz_2\,z_1^{1-2\epsilon}\,z_2^{1-2\epsilon}\,z_3^{1-2\epsilon}(\theta_{12}^2)^{-\epsilon}(\theta_{13}^2)^{-\epsilon}\sin^{-2\epsilon}\phi\,,
\end{align}
for phase space coordinates formed from pairwise angles $\theta_{12}^2,\theta_{13}^2$, azimuthal angle $\phi$, and energy fractions $z_i$, with $z_1+z_2+z_3=1$ on a jet with total energy $E_J$.  By the law of cosines, the azimuthal angle $\phi$ is defined in terms of the pairwise angles as
\begin{align}
\theta_{23}^2 = \theta_{12}^2+\theta_{13}^2-2\theta_{12}\theta_{13}\cos\phi\,.
\end{align}
$|{\cal M}_{q\to \bar q'_1q'_2 q_3}|^2$ is the squared matrix element for this collinear splitting where \cite{Campbell:1997hg,Catani:1998nv}
\begin{align}\label{eq:matelqqq}
|{\cal M}_{q\to \bar q'_1q'_2 q_3}|^2 &= \left(
8\pi \mu_0^{2\epsilon}\alpha_s
\right)^2\frac{C_F T_R}{2 E_J^4}\frac{1}{z_1z_2\theta_{12}^2\left(z_1z_2\theta_{12}^2+z_1z_3\theta_{13}^2+z_2z_3\theta_{23}^2\right)}\\
&\hspace{0.5cm}\times\Bigg[
-\frac{z_1z_2\theta^2_{12}}{z_1z_2\theta_{12}^2+z_1z_3\theta_{13}^2+z_2z_3\theta_{23}^2}\left(
\frac{2z_3}{1-z_3}\frac{\theta_{23}^2-\theta_{13}^2}{\theta_{12}^2}+\frac{z_1-z_2}{z_1+z_2}
\right)^2\nonumber\\
&\hspace{1cm}+\frac{4z_3+(z_1-z_2)^2}{z_1+z_2}+(1-2\epsilon)\left(
z_1+z_2-\frac{z_1z_2\theta_{12}^2}{z_1z_2\theta_{12}^2+z_1z_3\theta_{13}^2+z_2z_3\theta_{23}^2}
\right)
\Bigg]\,.\nonumber
\end{align}
Finally, $\Theta_\text{WTA}$ is the WTA flavor measurement function of the phase space variables that depends on the specific WTA flavor $f$ and scale $k_\perp^2$.  

At this order, the WTA flavor constraints are very complicated, so we can't necessarily hope to evaluate the jet function in closed, analytic form.  Therefore, we will restrict calculations to leading-logarithmic accuracy, at which splittings are described by strongly-ordered emissions.  In the hierarchical collinear limit of the $q'\bar q'$ pair, and averaged over their relative azimuthal angle, the squared matrix element is
\begin{align}
|{\cal M}_{q\to \bar q'_1q'_2 q_3}|^2 &\to \left(
8\pi \mu_0^{2\epsilon}\alpha_s
\right)^2\frac{C_F T_R}{2 E_J^4}\frac{1}{z_1z_2z_3(1-z_3)\theta_{12}^2\theta_{13}^2}\\
&\hspace{4cm}\times\left[
-\frac{8z_1z_2z_3}{(1-z_3)^3}+\frac{4z_3+(z_1-z_2)^2}{1-z_3}+(1-2\epsilon)\left(
1-z_3
\right)
\right]\,.\nonumber
\end{align}
We leave a thorough calculation of the jet function to future work, and its consequence on higher logarithmic accuracy WTA flavor evolution. 

\subsection{UV to IR Quark Flavor Change}

We will start our analysis here by studying contributions to the quark flavor changing jet function $J^{(2)}_{q\to q'}$ with $q'\neq q$, to validate the structure of the factorization \theorem~that is dependent on the UV renormalization scale $\mu^2$.

\subsubsection{Calculation from Evolution Equations}

Our first step will be to calculate the jet function $J_{q\to q'}(\mu^2,k_\perp^2)$ from the explicit solution to the evolution equations, \Eq{eq:gensol2loop}.  Because the quark flavors are distinct, $q\neq q'$, there are no contributions to the jet function at either ${\cal O}(\alpha_s^0)$ or ${\cal O}(\alpha_s^1)$.  The leading contributions start at ${\cal O}(\alpha_s^2)$ and can be expressed as
\begin{align}
J_{q\to q'}(\mu^2,k_\perp^2) &= \left(\frac{\alpha_s}{2\pi}\right)^2\left(
\frac{1}{2}\sum_k \gamma_{q\to k}^{(1)}\gamma_{k\to q'}^{(1)}\, \log^2\frac{\mu^2}{k_\perp^2}+\left(
\sum_k \gamma_{q\to k}^{(1)}c_{k\to q'}^{(1)}+\gamma_{q\to q'}^{(2)}
\right)\log\frac{\mu^2}{k_\perp^2}+c_{q\to q'}^{(2)}
\right)\nonumber\\
&
\hspace{1cm}+{\cal O}(\alpha_s^3)\,.
\end{align}
By the structure of the one-loop jet functions, the only possible intermediate particle $k$ that can connect two quarks of differing flavor is a gluon, and so the jet function to this order reduces to
\begin{align}
J_{q\to q'}(\mu^2,k_\perp^2) &= \left(\frac{\alpha_s}{2\pi}\right)^2\left(
\frac{1}{2} \gamma_{q\to g}^{(1)}\gamma_{g\to q'}^{(1)}\, \log^2\frac{\mu^2}{k_\perp^2}+\left(
 \gamma_{q\to g}^{(1)}c_{g\to q'}^{(1)}+\gamma_{q\to q'}^{(2)}
\right)\log\frac{\mu^2}{k_\perp^2}+c_{q\to q'}^{(2)}
\right)\nonumber\\
&
\hspace{1cm}+{\cal O}(\alpha_s^3)\,.
\end{align}
Plugging in the corresponding one-loop results summarized in \Sec{sec:summoneloop}, this jet function is
\begin{align}\label{eq:twoloop_renormsol}
J_{q\to q'}(\mu^2,k_\perp^2) &=\left(
\frac{\alpha_s}{2\pi}
\right)^2\left[
 C_F T_R \left(-\frac{5}{48}+\frac{1}{3}\log 2\right)\log^2\frac{\mu^2}{k_\perp^2}\right.\\
 &
 \hspace{2cm}\left.+\left(C_F T_R\left(
-\frac{5}{8}+2\log 2
\right)\left(
\frac{17}{36}-\frac{2}{3}\log 2
\right)+\gamma^{(2)}_{q\to q'}\right)\log\frac{\mu^2}{k_\perp^2}+c_{q\to q'}^{(2)}
 \right]\nonumber\\
&
\hspace{1cm}+{\cal O}(\alpha_s^3)\,.\nonumber
\end{align}
We are still missing the two-loop anomalous dimension $\gamma^{(2)}_{q\to q'}$ and constant $c_{q\to q'}^{(2)}$ for complete knowledge of the jet function through this order.  However, we will next validate that the leading-logarithmic term is correctly reproduced with the renormalization prescription which then correspondingly enables extraction of these values from the explicit calculation of the bare two-loop jet function.

\subsubsection{Calculation from Explicit Renormalization}

Using the general renormalization prescription of the jet function from \Sec{sec:jetrenorm}, we can evaluate the two-loop renormalized jet function for the $q\to q'$ transition via
\begin{align}
J_{q\to q'}^\text{(bare)} =  J_{q\to q'}^\text{(2,ren)}+Z_{q\to g}^{(1)} J_{g\to q'}^\text{(1,ren)}+Z_{q\to q'}^{(2)}+{\cal O}(\alpha_s^3)\,.
\end{align}
We will focus here on calculation of the leading logarithmic terms at ${\cal O}(\alpha_s^2)$ in this jet function, and so we only need to calculate each term in this relationship to leading logarithmic order.  The leading logarithmic contributions to the product of the one-loop renormalized jet function and the renormalization factor are
\begin{align}
Z_{q\to g}^{(1)} J_{g\to q'}^\text{(1,ren)} &\supset \left(\frac{\alpha_s}{2\pi}\right)^2 \frac{C_F T_R}{3}\left(
-\frac{5}{8}+2\log 2\right)\left(
\frac{\mu_0^2}{\mu^2}
\right)^\epsilon\frac{1}{\epsilon^2}\left[
 \left(
\frac{\mu_0^2}{k_\perp^2}
\right)^\epsilon - \left(
\frac{\mu_0^2}{\mu^2}
\right)^\epsilon
\right]
\\
&=\left(\frac{\alpha_s}{2\pi}\right)^2 C_F T_R \left[
\frac{1}{\epsilon}\left(-\frac{5}{24}+\frac{2}{3}\log 2\right)\log\frac{\mu^2}{k_\perp^2}+\left(-\frac{5}{12}+\frac{4}{3}\log 2\right)\log\frac{\mu_0^2}{\mu^2}\log\frac{\mu^2}{k_\perp^2}\right.\nonumber\\
&\hspace{5cm}
\left.
+\left(-\frac{5}{48}+\frac{1}{3}\log 2\right)\log^2\frac{\mu^2}{k_\perp^2}\right]\,.\nonumber
\end{align}

Next, we need to evaluate the bare jet function as described by the master formula of \Eq{eq:twoloopjetmaster}.  We have already presented the expression for the differential phase space and squared matrix element relevant for this transition; now, we need to construct the measurement function $\Theta_\text{WTA}$ that establishes the quark $q'$ as the WTA axis.  Because there are no flavorless particles in the $q\to \bar q_1' q_2'q_3 $ final state, there cannot be any particles within a relative transverse momentum scale $k_\perp^2$ of the quark $q'$.  The WTA axis is defined through pairwise reclustering with the $k_T$ algorithm, so, for a jet with three final state particles, we must consider three possible clustering histories.  If $q'$ is initially clustered with either $\bar q'$ or $q$, we must require that the $q'$ is more energetic than its cluster partner and together they carry more than half of the jet's energy.  If instead $\bar q'$ and $q$ are clustered together first, the $q'$ quark alone must carry more than half of the jet's energy.

These considerations produce the complete measurement function for WTA flavor of $q'$ in the $q\to \bar q_1' q_2'q_3 $ collinear splitting to be
\begin{align}
\Theta_\text{WTA}&=\Theta\left(\min[z_1^2,z_3^2]\theta_{13}^2-z_1^2\theta_{12}^2\right)\Theta\left(\min[z_2^2,z_3^2]\theta_{23}^2-z_1^2\theta_{12}^2\right)\\
&\hspace{2cm}\times\Theta\left(\frac{1}{2}-z_3\right)\Theta(z_2-z_1)\Theta\left(z_1^2\theta_{12}^2 E_J^2-k_\perp^2\right)\nonumber\\
&
\hspace{1cm}+\Theta\left(\min[z_1^2,z_2^2]\theta_{12}^2-z_3^2\theta_{23}^2\right)\Theta\left(\min[z_1^2,z_3^2]\theta_{13}^2-z_3^2\theta_{23}^2\right)\nonumber\\
&\hspace{2cm}\times\Theta\left(\frac{1}{2}-z_1\right)\Theta(z_2-z_3)\Theta\left(z_3^2\theta_{23}^2 E_J^2-k_\perp^2\right)\nonumber\\
&
\hspace{1cm}+\Theta\left(z_1^2\theta_{12}^2-\min[z_1^2,z_3^2]\theta_{13}^2\right)\Theta\left(z_3^2\theta_{23}^2-\min[z_1^2,z_3^2]\theta_{13}^2\right)\Theta\left(z_2-\frac{1}{2}\right)\nonumber\\
&
\hspace{2cm}
\times\left[
\Theta(z_1-z_3)\Theta\left((1-z_2)^2\theta_{12}^2E_J^2-k_\perp^2\right)+\Theta(z_3-z_1)\Theta\left((1-z_2)^2\theta_{23}^2 E_J^2-k_\perp^2\right)
\right]\,.
\nonumber
\end{align}
In these expressions, because we utilize the WTA recombination scheme, the relative transverse momentum from a reclustered pair of particles is measured with respect to the harder of the two in the cluster.  One thing to note for this measurement function is that its zero-bin explicitly vanishes in the limit that $z_1,z_2\sim k_\perp\to 0$.

Restricting to validating the leading-logarithms of the jet function, the measurement function simplifies significantly.  At leading logarithmic accuracy, the $\bar q'_1 q'_2$ pair are parametrically more collinear than their angle to the final-state $q_3$, and so we can expand the measurement function in the limit where $\theta_{12}^2 \ll \theta_{13}^2\sim \theta_{23}^2$:
\begin{align}
\Theta_\text{WTA}&\to \Theta(\theta_{13}^2 - \theta_{12}^2)\Theta\left(\frac{1}{2}-z_3\right)\Theta(z_2-z_1)\Theta\left(\theta_{12}^2-\frac{k_\perp^2}{E_J^2}\right)\,.
\end{align}
Because we require that $q'_2$ is relatively hard, there are no divergent soft limits for this flavor configuration, so just to extract the leading logarithms, we can ignore energy fraction modifications to angles, as they will only vary the argument of the logs by an order-1 amount, and are therefore subleading.

With these strongly-ordered collinear results, we can then integrate over the angles $\theta_{13}^2$ and $\theta_{12}^2$ and extract the leading divergent terms.  We find
\begin{align}
J_{q\to q'}^{(2,\text{bare})} &= \int d\Phi_3\,|{\cal M}|^2\,\Theta_\text{WTA}\\
&\supset \left(\frac{\alpha_s}{2\pi}\right)^2\frac{C_F T_R}{4}\left(
\frac{\mu_0^2}{k_\perp^2}
\right)^{2\epsilon}\frac{1}{\epsilon^2}\int dz_1\, dz_2\, \left[-\frac{8z_1z_2z_3}{(1-z_3)^4}
+\frac{4z_3+(z_1-z_2)^2}{(1-z_3)^2}+1\right]\nonumber\\
&\hspace{7cm}
\times \Theta(1-z_1-z_2)\Theta\left(\frac{1}{2}-z_3\right)\Theta(z_2-z_1)\,,\nonumber
\end{align}
where we note that $z_3=1-z_2-z_1$.  Performing the integral, we find
\begin{align}\label{eq:final}
J_{q\to q'}^{(2,\text{bare})} &\supset \left(\frac{\alpha_s}{2\pi}\right)^2\frac{C_F T_R}{4}\left(
\frac{\mu_0^2}{k_\perp^2}
\right)^{2\epsilon}\frac{1}{\epsilon^2}\left(
-\frac{5}{12}+\frac{4}{3}\log 2
\right)\\
&=\left(\frac{\alpha_s}{2\pi}\right)^2C_F T_R\left[\frac{1}{4\epsilon^2}\left(
-\frac{5}{12}+\frac{4}{3}\log 2
\right)+\frac{1}{\epsilon}\left(
-\frac{5}{24}+\frac{2}{3}\log 2
\right)\log\frac{\mu_0^2}{k_\perp^2}\right.\nonumber\\
&
\hspace{8cm}
\left.+\left(
-\frac{5}{24}+\frac{2}{3}\log 2
\right)\log^2\frac{\mu_0^2}{k_\perp^2}
\right]\,.\nonumber
\end{align}

Then, the difference of the bare jet function and the product of one-loop quantities is
\begin{align}
J_{q\to q'}^\text{(2,bare)} - Z_{q\to g}^{(1)} J_{g\to q'}^\text{(1,ren)}&= J_{q\to q'}^\text{(2,ren)}+Z_{q\to q'}^{(2)}\\
&\supset \left(\frac{\alpha_s}{2\pi}\right)^2C_F T_R\left[\frac{1}{4\epsilon^2}\left(
-\frac{5}{12}+\frac{4}{3}\log 2
\right)+\frac{1}{\epsilon}\left(
-\frac{5}{24}+\frac{2}{3}\log 2
\right)\log\frac{\mu_0^2}{\mu^2}
\right]\nonumber\\
&\hspace{1cm}
+\left(\frac{\alpha_s}{2\pi}\right)^2C_F T_R\left(
-\frac{5}{24}+\frac{2}{3}\log 2
\right)\left(\log^2\frac{\mu_0^2}{\mu^2}+\frac{1}{2}\log^2\frac{\mu^2}{k_\perp^2}\right)
\nonumber\,.
\end{align}
All remaining divergences can be absorbed into the two-loop renormalization factor, where 
\begin{align}
Z_{q\to q'}^{(2)} &= \left(\frac{\alpha_s}{2\pi}\right)^2C_F T_R\left(
-\frac{5}{12}+\frac{4}{3}\log 2
\right)\left(\frac{1}{4\epsilon^2}+\frac{1}{2\epsilon}\log\frac{\mu_0^2}{\mu^2}+\frac{1}{2}\log^2\frac{\mu_0^2}{\mu^2}\right)\\
&\to \left(\frac{\alpha_s}{2\pi}\right)^2C_F T_R\left(
-\frac{5}{12}+\frac{4}{3}\log 2
\right)\frac{1}{4\epsilon^2}\left(
\frac{\mu_0^2}{\mu^2}
\right)^{2\epsilon}\,.\nonumber
\end{align}
Then, the renormalized two-loop jet function for this flavor-changing process is
\begin{align}
 J_{q\to q'}^\text{(2,ren)}\supset \left(\frac{\alpha_s}{2\pi}\right)^2C_F T_R\left(
-\frac{5}{48}+\frac{1}{3}\log 2
\right)\log^2\frac{\mu^2}{k_\perp^2}\,,
\end{align}
which agrees exactly with the leading-logarithmic terms in the solution of the renormalization group equation to this order, Eq.~\ref{eq:twoloop_renormsol}.

\subsection{UV to IR Quark-Gluon Flavor Change in $C_Fn_f T_R$ Channel}\label{sec:as2calcs_qg}

The second WTA flavor configuration we consider is the transition $q\to g$ in the color channel of this squared matrix element.  This is especially interesting because this flavor restriction requires that two collinear emissions $q'\bar q'$ are unresolved, and we must sum over all possible quark flavors:
\begin{align}
|{\cal M}_{q\to  g_{12} q_3}|^2 \equiv \sum_{q'}|{\cal M}_{q\to \bar q'_1q'_2 q_3}|^2 = n_f|{\cal M}_{q\to \bar q'_1q'_2 q_3}|^2\,,
\end{align}
for $n_f$ massless quark species.

\subsubsection{Calculation from Evolution Equations}

To this order and in this color channel, the leading logarithmic terms from the solution of the evolution equations are expressed as
\begin{align}
J_{q\to g,n_fT_R}^{(2)}(\mu^2,k_\perp^2) &\supset\left(
\frac{\alpha_s}{2\pi}
\right)^2
  \frac{\gamma^{(1)}_{q\to g}\, \gamma^{(1)}_{g\to g}+\beta_0\,\gamma^{(1)}_{q\to g}}{2}\,\log^2\frac{\mu^2}{k_\perp^2}\,.
\end{align}
Note the presence of the $\beta$-function coefficient, because of its $n_fT_R$ term.  Plugging in the values of the anomalous dimensions, we have
\begin{align}\label{eq:nftrjetevo}
J_{q\to g,n_fT_R}^{(2)}(\mu^2,k_\perp^2) &\supset-\left(
\frac{\alpha_s}{2\pi}
\right)^2
C_F n_f T_R\left(
-\frac{5}{12}+\frac{4}{3}\log 2
\right) \log^2\frac{\mu^2}{k_\perp^2}\,.
\end{align}

\subsubsection{Calculation from Explicit Renormalization}

To explicitly calculate this jet function from the collinear splitting matrix elements, we note that the complete measurement function for the transition $q\to g$ here is
\begin{align}\label{eq:q2gcfnf}
\Theta_\text{WTA} &= \Theta\left(\min[z_1^2,z_3^2]\theta_{13}^2-\min[z_1^2,z_2^2]\theta_{12}^2\right)\Theta\left(\min[z_2^2,z_3^2]\theta_{23}^2-\min[z_1^2,z_2^2]\theta_{12}^2\right)\\
&\hspace{2cm}\times\Theta\left(\frac{1}{2}-z_3\right)\Theta\left(k_\perp^2-\min[z_1^2,z_2^2]\theta_{12}^2 E_J^2\right)\nonumber\\
&\hspace{1cm}\times
\left[
\Theta(z_1-z_2)\Theta\left(z_3^2\theta_{13}^2 E_J^2-k_\perp^2\right)+\Theta(z_2-z_1)\Theta\left(z_3^2\theta_{23}^2 E_J^2-k_\perp^2\right)
\right]\nonumber\,.
\end{align}
We require that the pair $\bar q_1' q_2'$ are unresolved, and have a relative transverse momentum less than $k_\perp^2$.  However, their relative transverse momentum to $q_3$ must be larger than $k_\perp^2$, to ensure that they have a flavor distinct from the initiating quark.  Focusing on the leading-logarithmic limit, we take the collinear limit in which $\bar q_1' q_2'$ are strongly-ordered where the measurement function then becomes
\begin{align}
\Theta_\text{WTA} &\to \Theta\left(\frac{1}{2}-z_3\right)\Theta\left(k_\perp^2-\theta_{12}^2 E_J^2\right)\Theta\left(\theta_{13}^2 E_J^2-k_\perp^2\right)\,.
\end{align}
Again, we can ignore overall scalings by the energy fractions because all must be order-1.  This correspondingly means that the zero-bin phase space constraints of \Eq{eq:q2gcfnf} vanish as well, taking the scaling $z_1,z_2\sim k_\perp\to 0$.

Using this measurement function, the bare jet function for this flavor transition is
\begin{align}
J_{q\to g,n_fT_R}^{(2,\text{bare})} &= \int d\Phi_3|{\cal M}|^2\Theta_\text{WTA}\\
&\supset -\left(\frac{\alpha_s}{2\pi}\right)^2\frac{C_F n_fT_R}{2}\left(
\frac{\mu_0^2}{k_\perp^2}
\right)^{2\epsilon}\frac{1}{\epsilon^2}\int dz_1\, dz_2\, \left[-\frac{8z_1z_2z_3}{(1-z_3)^4}
+\frac{4z_3+(z_1-z_2)^2}{(1-z_3)^2}+1\right]\nonumber\\
&\hspace{9cm}
\times \Theta\left(\frac{1}{2}-z_3\right)\nonumber\\
&=-\left(\frac{\alpha_s}{2\pi}\right)^2C_F n_fT_R\left(
\frac{\mu_0^2}{k_\perp^2}
\right)^{2\epsilon}\frac{1}{\epsilon^2}\left(
-\frac{5}{12}+\frac{4}{3}\log 2
\right)\,.\nonumber
\end{align}
The renormalization prescription relates this bare jet function to the renormalized jet function as
\begin{align}
J_{q\to g}^\text{(2,bare)} =   J_{q\to g}^\text{(2,ren)}+Z_{q\to g}^{(1)} J_{g\to g}^\text{(1,ren)}+\left(Z_{q\to q}^{(1)}+\frac{\alpha_s}{2\pi}\beta_0 \frac{1}{\epsilon}\left(
\frac{\mu_0^2}{\mu^2}
\right)^\epsilon\right) J_{q\to g}^\text{(1,ren)}+Z_{q\to g}^{(2)}\,.
\end{align}
Note that the color factor of the $Z_{q\to q}^{(1)} J_{q\to g}^\text{(1,ren)}$ term is $C_F^2$, and so doesn't contribute to the case at hand.  The explicit $\beta$-function term is needed to modify the scale at which the coupling is evaluated in $J_{q\to g}^\text{(1,ren)}$ from $\mu^2$ to $\mu_0^2$.  Therefore, the renormalized jet function is
\begin{align}
  J_{q\to g,n_fT_R}^\text{(2,ren)}=J_{q\to g,n_fT_R}^\text{(2,bare)} -Z_{q\to g}^{(1)} J_{g\to g}^\text{(1,ren)}-\frac{\alpha_s}{2\pi}\beta_0 \frac{1}{\epsilon}\left(
\frac{\mu_0^2}{\mu^2}
\right)^\epsilon J_{q\to g}^\text{(1,ren)}-Z_{q\to g,n_fT_R}^{(2)}\,.
\end{align}

The difference of the bare and one-loop renormalized factors is
\begin{align}
&J_{q\to g,n_fT_R}^\text{(2,bare)} -Z_{q\to g}^{(1)} J_{g\to g}^\text{(1,ren)}-\frac{\alpha_s}{2\pi}\beta_0 \frac{1}{\epsilon}\left(
\frac{\mu_0^2}{\mu^2}
\right)^\epsilon J_{q\to g}^\text{(1,ren)} \\
&\hspace{2cm}\supset -\left(\frac{\alpha_s}{2\pi}\right)^2C_F n_fT_R\left(
-\frac{5}{12}+\frac{4}{3}\log 2
\right)\frac{1}{\epsilon^2}\left[\left(
\frac{\mu_0^2}{k_\perp^2}
\right)^{2\epsilon}+2\left(
\frac{\mu_0^2}{\mu^2}
\right)^{2\epsilon}-2\left(
\frac{\mu_0^4}{\mu^2k_\perp^2}
\right)^{\epsilon}\right]\nonumber\\
&=-\left(\frac{\alpha_s}{2\pi}\right)^2C_F n_fT_R\left(
-\frac{5}{12}+\frac{4}{3}\log 2
\right)\frac{1}{\epsilon^2}\left[
1+2\epsilon\log\frac{\mu_0^2}{\mu^2}+2\epsilon^2\log^2\frac{\mu_0^2}{\mu^2}+\epsilon^2\log^2\frac{\mu^2}{k_\perp^2}+{\cal O}(\epsilon^3)
\right]\,.\nonumber
\end{align}
Then, all residual divergences can be absorbed by setting the two-loop renormalization factor to
\begin{align}
Z_{q\to g,n_fT_R}^{(2)}=-\left(\frac{\alpha_s}{2\pi}\right)^2C_F n_fT_R\left(
-\frac{5}{12}+\frac{4}{3}\log 2
\right)\frac{1}{\epsilon^2}\left(
\frac{\mu_0^2}{\mu^2}
\right)^{2\epsilon}\,,
\end{align}
so that the leading-logarithmic terms of the renormalized jet function are
\begin{align}
  J_{q\to g,n_fT_R}^\text{(2,ren)}\supset -\left(\frac{\alpha_s}{2\pi}\right)^2C_F n_fT_R\left(
-\frac{5}{12}+\frac{4}{3}\log 2
\right)\log^2\frac{\mu^2}{k_\perp^2}\,,
\end{align}
which agrees exactly with \Eq{eq:nftrjetevo}.

\subsection{UV to IR Quark Flavor Conservation in $C_Fn_f T_R$ Channel}

The final partonic WTA flavor in this splitting is the flavor-diagonal transition, $q\to q$.  Now, both resolved and unresolved splittings contribute to this transition and further, the zero-bin will be non-trivial.  As with the transition $q\to g$, we will need to sum over all possible flavors of the pair $q'\bar q'$ in the matrix element.

\subsubsection{Calculation from Evolution Equations}

From the fixed-order expansion of the solution to the evolution equations, the possible leading logarithmic contributions to $q\to q$ transition take the form
\begin{align}
J_{q\to q}^{(2)}(\mu^2,k_\perp^2) &\supset\left(
\frac{\alpha_s}{2\pi}
\right)^2
  \frac{\gamma^{(1)}_{q\to q}\, \gamma^{(1)}_{q\to q}+\gamma^{(1)}_{q\to g}\, \gamma^{(1)}_{g\to q}+\beta_0\,\gamma^{(1)}_{q\to q}}{2}\,\log^2\frac{\mu^2}{k_\perp^2}\,.
\end{align}
Note that neither $\gamma^{(1)}_{q\to q}\, \gamma^{(1)}_{q\to q}$ nor $\gamma^{(1)}_{q\to g}\, \gamma^{(1)}_{g\to q}$ are proportional to $n_f$; only the $\beta_0$ term is.  Therefore, we have
\begin{align}\label{eq:q2qcfnf}
J_{q\to q,C_Fn_F T_R}^{(2)}(\mu^2,k_\perp^2) &\supset\left(
\frac{\alpha_s}{2\pi}
\right)^2
  \frac{\beta_0\,\gamma^{(1)}_{q\to q}}{2}\,\log^2\frac{\mu^2}{k_\perp^2}\\
  &=\left(\frac{\alpha_s}{2\pi}\right)^2C_F n_fT_R\left(
-\frac{5}{24}+\frac{2}{3}\log 2
\right)\log^2\frac{\mu^2}{k_\perp^2}\nonumber
\end{align}

\subsubsection{Calculation from Explicit Renormalization}

Now, we move to explicit calculation of the renormalized jet function from its bare version.  The measurement function for WTA $q$ flavor is a rather large mess; namely
\begin{align}
\Theta_\text{WTA} &= \Theta\left(z_1^2\theta_{13}^2-\min[z_1^2,z_2^2]\theta_{12}^2\right)\Theta\left(z_2^2\theta_{23}^2-\min[z_1^2,z_2^2]\theta_{12}^2\right)\\
&\hspace{1cm}\times\Theta\left(z_3-\frac{1}{2}\right)
\left[
\Theta(z_1-z_2)\Theta\left(z_1^2\theta_{13}^2 E_J^2-k_\perp^2\right)+\Theta(z_2-z_1)\Theta\left(z_2^2\theta_{23}^2 E_J^2-k_\perp^2\right)
\right]\nonumber\\
&+ \Theta\left(\min[z_1^2,z_3^2]\theta_{13}^2-\min[z_1^2,z_2^2]\theta_{12}^2\right)\Theta\left(\min[z_2^2,z_3^2]\theta_{23}^2-\min[z_1^2,z_2^2]\theta_{12}^2\right)\nonumber\\
&\hspace{1cm}\times
\left[
\Theta(z_1-z_2)\Theta\left(k_\perp^2-\min[z_1^2,z_3^2]\theta_{13}^2 E_J^2\right)+\Theta(z_2-z_1)\Theta\left(k_\perp^2-\min[z_2^2,z_3^2]\theta_{23}^2 E_J^2\right)
\right]\nonumber\\
&+\Theta\left(\min[z_1^2,z_2^2]\theta_{12}^2-\min[z_1^2,z_3^2]\theta_{13}^2\right)\Theta\left(\min[z_2^2,z_3^2]\theta_{23}^2-\min[z_1^2,z_3^2]\theta_{13}^2\right)\nonumber\\
&\hspace{1cm}\times\Theta\left(\min[z_1^2,z_3^2]\theta_{13}^2E_J^2-k_\perp^2\right)\Theta\left(z_1+z_3-\frac{1}{2}\right)\Theta(z_3-z_1)\nonumber\\
&+\Theta\left(\min[z_1^2,z_2^2]\theta_{12}^2-\min[z_2^2,z_3^2]\theta_{23}^2\right)\Theta\left(\min[z_1^2,z_3^2]\theta_{13}^2-\min[z_2^2,z_3^2]\theta_{23}^2\right)\nonumber\\
&\hspace{1cm}\times\Theta\left(\min[z_2^2,z_3^2]\theta_{23}^2E_J^2-k_\perp^2\right)\Theta\left(z_2+z_3-\frac{1}{2}\right)\Theta(z_3-z_2)\nonumber\,.
\end{align}
In the first two lines, the $\bar q_1'q_2'$ pair is clustered first, but resolved from the $q_3$ and relatively soft; in the third and forth lines, all particles are within $k_\perp$ of the $q$ and so there is no energy constraint imposed; on the fifth and sixth (seventh and eighth) lines, $\bar q_1'$ and $q_3$ ($q_2'$ and $q_3$) are clustered first but resolved, separated beyond a relative transverse momentum of $k_\perp^2$.  Unlike the previous two flavor transitions considered, there is a non-trivial zero-bin of the phase space constraints from taking $z_1,z_2\sim k_\perp\to 0$, where
\begin{align}\label{eq:soft_phase_space}
\Theta_\text{WTA}^{(0)} &=\Theta\left(\min[z_1^2,z_2^2]\theta_{12}^2-z_1^2\theta_{13}^2\right)\Theta\left(z_2^2\theta_{23}^2-z_1^2\theta_{13}^2\right)\Theta\left(z_1^2\theta_{13}^2E_J^2-k_\perp^2\right)\\
&\hspace{1cm}+\Theta\left(\min[z_1^2,z_2^2]\theta_{12}^2-z_2^2\theta_{23}^2\right)\Theta\left(z_1^2\theta_{13}^2-z_2^2\theta_{23}^2\right)\Theta\left(z_2^2\theta_{23}^2E_J^2-k_\perp^2\right)\nonumber\,.
\end{align}
Here, we have ignored constraints that produce manifestly scaleless integrals. While this zero-bin will not be needed at leading logarithm, it is critical to check that no soft contribution affects the extraction of the two-loop anomalous dimension of the jet function. We show in App. \ref{app:two_loop_soft} this contribution is indeed scaleless and therefore vanishes after appropriate rescalings of the variables. 

For our leading-logarithmic calculation, the collinear limit of the measurement function is
\begin{align}
\Theta_\text{WTA} 
&\to\Theta\left(\min[z_1^2,z_3^2]\theta_{13}^2-\min[z_1^2,z_2^2]\theta_{12}^2\right)\Theta\left(\min[z_2^2,z_3^2]\theta_{13}^2-\min[z_1^2,z_2^2]\theta_{12}^2\right)\\
&\hspace{-1cm}\times
\Theta\left(\frac{1}{2}-z_3\right)\left[
\Theta(z_1-z_2)\Theta\left(k_\perp^2-\min[z_1^2,z_3^2]\theta_{13}^2 E_J^2\right)+\Theta(z_2-z_1)\Theta\left(k_\perp^2-\min[z_2^2,z_3^2]\theta_{13}^2 E_J^2\right)
\right]\nonumber\,,
\end{align}
where we have ignored scaleless phase space constraints.  Note that these constraints then eliminate the soft divergence $z_1,z_2\to 0$, and so, for extracting the leading-logarithmic prediction all energy fractions are order-1, and therefore their specific values only contribute at next-to-leading logarithmic order.  Then, at leading logarithmic accuracy, this phase space constraint can be expressed as
\begin{align}
\Theta_\text{WTA} &\to\Theta\left(\theta_{13}^2-\theta_{12}^2\right)
\Theta\left(\frac{1}{2}-z_3\right)\Theta\left(k_\perp^2-\theta_{13}^2 E_J^2\right)\,,
\end{align}

Using this measurement function, the bare jet function for this flavor transition is
\begin{align}
J_{q\to q,n_fT_R}^{(2,\text{bare})} &= \int d\Phi_3|{\cal M}|^2\Theta_\text{WTA}\\
&\supset \left(\frac{\alpha_s}{2\pi}\right)^2\frac{C_F n_fT_R}{4}\left(
\frac{\mu_0^2}{k_\perp^2}
\right)^{2\epsilon}\frac{1}{\epsilon^2}\int dz_1\, dz_2\, \left[-\frac{8z_1z_2z_3}{(1-z_3)^4}
+\frac{4z_3+(z_1-z_2)^2}{(1-z_3)^2}+1\right]\nonumber\\
&\hspace{9cm}
\times \Theta\left(\frac{1}{2}-z_3\right)\nonumber\\
&=\left(\frac{\alpha_s}{2\pi}\right)^2C_F n_fT_R\left(
\frac{\mu_0^2}{k_\perp^2}
\right)^{2\epsilon}\frac{1}{\epsilon^2}\left(
-\frac{5}{24}+\frac{2}{3}\log 2
\right)\,.\nonumber
\end{align}
The renormalization prescription relates this bare jet function to the renormalized jet function as
\begin{align}
J_{q\to q}^\text{(2,bare)} =   J_{q\to q}^\text{(2,ren)}+Z_{q\to g}^{(1)} J_{g\to q}^\text{(1,ren)}+\left(Z_{q\to q}^{(1)}+\frac{\alpha_s}{2\pi}\beta_0 \frac{1}{\epsilon}\left(
\frac{\mu_0^2}{\mu^2}
\right)^\epsilon\right) J_{q\to q}^\text{(1,ren)}+Z_{q\to q}^{(2)}\,.
\end{align}
Of the terms involving one-loop renormalization, only the $\beta$-function term is proportional to $n_f$, and its difference with the bare jet function is
\begin{align}
&J_{q\to q}^\text{(2,bare)}  - \frac{\alpha_s}{2\pi}\beta_0 \frac{1}{\epsilon}\left(
\frac{\mu_0^2}{\mu^2}
\right)^\epsilon J_{q\to q}^\text{(1,ren)} \\
&\hspace{1cm}\supset\left(\frac{\alpha_s}{2\pi}\right)^2C_Fn_fT_R \frac{1}{\epsilon^2}\left(
-\frac{5}{24}+\frac{2}{3}\log 2\right)\left[
\left(
\frac{\mu_0^2}{k_\perp^2}
\right)^{2\epsilon}+2\left(
\frac{\mu_0^2}{\mu^2}
\right)^{2\epsilon}-2\left(
\frac{\mu_0^4}{\mu^2k_\perp^2}
\right)^\epsilon
\right]\nonumber\\
&=\left(\frac{\alpha_s}{2\pi}\right)^2C_F n_fT_R\left(
-\frac{5}{24}+\frac{2}{3}\log 2
\right)\frac{1}{\epsilon^2}\left[
1+2\epsilon\log\frac{\mu_0^2}{\mu^2}+2\epsilon^2\log^2\frac{\mu_0^2}{\mu^2}+\epsilon^2\log^2\frac{\mu^2}{k_\perp^2}+{\cal O}(\epsilon^3)
\right]\,.\nonumber
\end{align}
All residual divergences can be absorbed by setting the two-loop renormalization factor to
\begin{align}
Z_{q\to q,n_fT_R}^{(2)}=\left(\frac{\alpha_s}{2\pi}\right)^2C_F n_fT_R\left(
-\frac{5}{24}+\frac{2}{3}\log 2
\right)\frac{1}{\epsilon^2}\left(
\frac{\mu_0^2}{\mu^2}
\right)^{2\epsilon}\,,
\end{align}
so that the leading-logarithmic terms of the renormalized jet function are
\begin{align}
  J_{q\to q,n_fT_R}^\text{(2,ren)}\supset \left(\frac{\alpha_s}{2\pi}\right)^2C_F n_fT_R\left(
-\frac{5}{24}+\frac{2}{3}\log 2
\right)\log^2\frac{\mu^2}{k_\perp^2}\,,
\end{align}
which agrees exactly with \Eq{eq:q2qcfnf}.

\subsection{Non-Partonic Flavor: $q\to (qq')$}

Finally, with this matrix element, we must also comment on the possibility that net non-partonic flavor is contained with a region of transverse momentum $k_\perp^2$ of the WTA axis.  There are two possible configurations, but we will just focus on the transition where $q\to (qq')$ and $q'\neq \bar q$, where $(qq')$ means that the quarks $q$ and $q'$ compose the WTA region.  The measurement function for this configuration is
\begin{align}
\Theta_\text{WTA} &= \Theta\left(\min[z_1^2,z_2^2]\theta_{12}^2-\min[z_2^2,z_3^2]\theta_{23}^2\right)\Theta\left(\min[z_1^2,z_3^2]\theta_{13}^2-\min[z_2^2,z_3^2]\theta_{23}^2\right)\\
&\hspace{1cm}\times\Theta\left(k_\perp^2-\min[z_2^2,z_3^2]\theta_{23}^2E_J^2\right)\Theta\left(z_2+z_3-\frac{1}{2}\right)\nonumber\\
&\hspace{1cm}\times\left[
\Theta(z_2-z_3)\Theta\left(z_1^2\theta_{12}^2E_J^2-k_\perp^2\right)+\Theta(z_3-z_2)\Theta\left(z_1^2\theta_{13}^2E_J^2-k_\perp^2\right)
\right]\nonumber
\end{align}
Note that the quark $\bar q_1'$ must be resolved, so this measurement function forbids the limit in which the pair $\bar q_1'q_2'$ become collinear, $\theta_{12}\ll \theta_{13},\theta_{23}$.  Thus, only the overall collinear limit, $\theta_{12}\sim \theta_{13}\sim \theta_{23}\sim k_\perp\to 0$, can possibly contribute.

However, there is a non-trivial soft limit of the phase space, $z_1,z_2\sim k_\perp \to 0$.  The zero-bin measurement function is
\begin{align}\label{eq:soft_phase_space_II}
\Theta_\text{WTA}^{(0)} &= \Theta\left(\min[z_1^2,z_2^2]\theta_{12}^2-z_2^2\theta_{23}^2\right)\Theta\left(z_1^2\theta_{13}^2-z_2^2\theta_{23}^2\right)\Theta\left(k_\perp^2-z_2^2\theta_{23}^2E_J^2\right)\\
&\hspace{1cm}\times\Theta\left(z_1^2\theta_{13}^2E_J^2-k_\perp^2\right)
\nonumber\,.
\end{align}
Following the analysis given in App. \ref{app:two_loop_soft}, we actually find this contribution is once again scaleless.

This is further a strong validation of the factorization \theorem~because a non-scaleless zero-bin would indicate that there would be non-trivial production of non-partonic WTA flavor from one-loop renormalization.  However, the fact that $J_{q\to(qq')}^{(2)}$ is non-zero but finite means that at higher orders, there will be collinear divergences and in general, $J_{q\to(qq')}$ will need renormalization.  This is entirely consistent with the factorization \theorem, because jet functions are only renormalized ``on the left''; that is, at the scale $\mu^2$, evolved up from the measurement scale $k_\perp^2$.  The jet flavor in the UV as produced from the hard function must be partonic, but can become non-partonic in the IR because of the finite size of $k_\perp^2$.  In particular, the renormalization group evolution of this non-partonic flavor is identical to that presented in \Sec{sec:qevoallords} for WTA quark flavor $q$, with the replacement in the jet functions of $j\to q$ to $j\to (qq')$.  The anomalous dimension matrix would be identical.

\section{Numerical Analysis}\label{sec:numerics}

In this section, we will use the results derived above and present predictions for the WTA flavor evolution at leading-logarithmic (LL) and leading-logarithmic prime (LL$'$) accuracy.  Here, LL$'$ accuracy means that one-loop low-scale constants are included in the hard and jet functions, which are formally only needed for next-to-leading logarithmic accuracy (NLL).  This will enable a study of the sensitivity of the scale dependence as well as improved in precision as higher-orders are included.  Nevertheless, a complete study at NLL is necessary, especially to match to next-to-next-to-leading order predictions.

What we will plot is the cross-section prediction from the factorization \theorem
\begin{align}
\sigma_{e^+e^-\to f_L f_R} &=\sum_{j_L,\,j_R} H_{e^+e^-\to j_L j_R}(Q^2,\mu^2)\, J_{j_L\to f_L}(\mu^2,k_\perp^2)\, J_{j_R\to f_R}(\mu^2,k_\perp^2)\,,
\end{align}
as a function of the center-of-mass collision energy $Q$, with a fixed value for the measurement scale, $k_\perp = 2$ GeV.  This measurement scale is comparable to the scale of the end of the parton shower and at which non-perturbative effects begin to become important.  However, 2 GeV is still sufficiently perturbative that we can study the effect of variations of this scale on the results.  By construction, the factorization \theorem~is independent of the renormalization scale $\mu^2$, so we can conveniently set it to $\mu^2 = Q^2$, for which the hard function is just evaluated at its natural scale and does not evolve:
\begin{align}
\sigma_{e^+e^-\to f_L f_R} &=\sum_{j_L,\,j_R} H_{e^+e^-\to j_L j_R}(Q^2,Q^2)\, J_{j_L\to f_L}(Q^2,k_\perp^2)\, J_{j_R\to f_R}(Q^2,k_\perp^2)\,.
\end{align}
While there are numerous possible WTA flavor combinations, here we will restrict ourselves to just displaying three results for brevity.  Nevertheless, all possible flavor configurations through LL$'$ accuracy can be predicted with the results summarized in the appendices.

Another thing to note is that, because the hard function is only evaluated at one-loop and does not evolve, this dramatically reduces the sum over intermediate parton flavors.  At one-loop, the only possible flavor change is of one of the initial quarks into a gluon, and so the WTA flavor cross section can be expressed as
\begin{align}
\sigma_{e^+e^-\to f_L f_R} &=\sum_{\text{flavor }q}\left[H_{e^+e^-\to q\bar q}(Q^2,Q^2)\, J_{q\to f_L}(Q^2,k_\perp^2)\, J_{\bar q\to f_R}(Q^2,k_\perp^2)\right.\\
&
\hspace{2cm}+H_{e^+e^-\to qg}(Q^2,Q^2)\, J_{q\to f_L}(Q^2,k_\perp^2)\, J_{g\to f_R}(Q^2,k_\perp^2)\nonumber\\
&\hspace{2cm}\left.+H_{e^+e^-\to g\bar q}(Q^2,Q^2)\, J_{g\to f_L}(Q^2,k_\perp^2)\, J_{\bar q\to f_R}(Q^2,k_\perp^2)\right]\,.\nonumber
\end{align}
Additionally, for simplicity, we will consider the natural scales of both of the jet functions to be identical and equal to the scale $\mu_J^2$, which is of the order of $k_\perp^2$.  This scale choice is not exhaustive nor the most general, but will demonstrate sensitivity to higher-order effects.

We first consider the case of quark flavor conservation, $\sigma_{e^+e^-\to q\bar q}$, isolating the term in the sum over intermediate flavors $q$ for which $q = f_L$.  This contribution to the WTA flavor cross section can be expressed as
\begin{align}
\sigma_{e^+e^-\to q\bar q} &\supset H_{e^+e^-\to q\bar q}(Q^2,Q^2)\, J_{q\to q}(Q^2,k_\perp^2)\, J_{\bar q\to \bar q}(Q^2,k_\perp^2)\\
&
\hspace{-1cm}+H_{e^+e^-\to qg}(Q^2,Q^2)\, J_{q\to q}(Q^2,k_\perp^2)\, J_{g\to \bar q}(Q^2,k_\perp^2)+H_{e^+e^-\to g\bar q}(Q^2,Q^2)\, J_{g\to q}(Q^2,k_\perp^2)\, J_{\bar q\to \bar q}(Q^2,k_\perp^2)\,.\nonumber
\end{align}
With the natural scale of the jet functions the same, then we note that this further simplifies, exploiting charge conjugation, 
\begin{align}
\sigma_{e^+e^-\to q\bar q} &\supset H_{e^+e^-\to q\bar q}(Q^2,Q^2) \left[J_{q\to q}(Q^2,k_\perp^2)\right]^2\\
&
\hspace{1cm}+2H_{e^+e^-\to qg}(Q^2,Q^2)\, J_{q\to q}(Q^2,k_\perp^2)\, J_{g\to q}(Q^2,k_\perp^2)\,.\nonumber
\end{align}

\begin{figure}[t!]
\begin{center}
\includegraphics[width=0.45\textwidth]{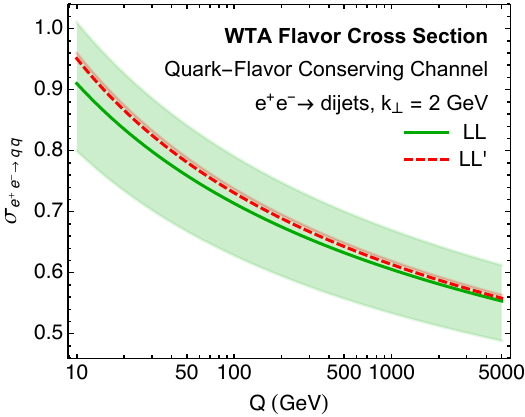}\ \ \ \includegraphics[width=0.45\textwidth]{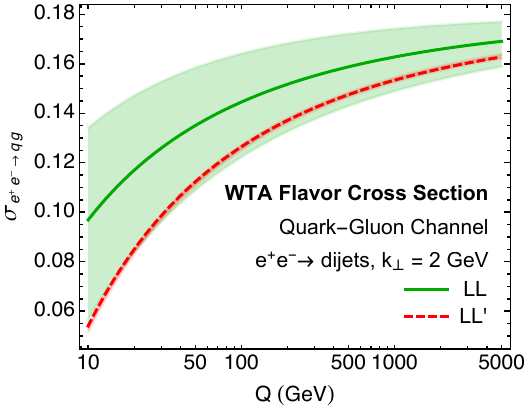}\\
\vspace{0.2cm}
\includegraphics[width=0.45\textwidth]{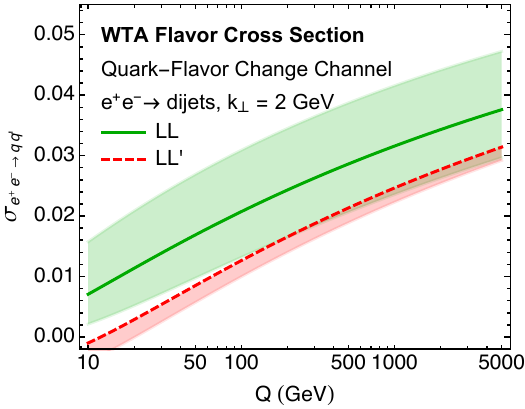}
\caption{\label{fig:eexsec}
Plots of the cross sections of the WTA flavor in each hemisphere of $e^+e^-\to$ hadrons events as a function of center-of-mass collision energy $Q$, for the transitions $e^+e^-\to q\bar q$ (upper left), $e^+e^-\to qg$ (upper right), and $e^+e^- \to qq'$ (bottom), inclusive over all $q'\neq q,\bar q$.  The measurement scale is set to $k_\perp = 2$ GeV and the number of active quarks is $n_f = 5$.  Uncertainty bands represent variation of the natural scale in the jet function by a factor of 2 about $k_\perp$.
}
\end{center}
\end{figure}

Next, we consider the case in which an initial anti-quark, for example, has transformed into a gluon, $\sigma_{e^+e^-\to qg}$, but the quark flavor $q$ is conserved.  We have
\begin{align}
\sigma_{e^+e^-\to qg} &\supset H_{e^+e^-\to q\bar q}(Q^2,Q^2)\, J_{q\to q}(Q^2,k_\perp^2)\, J_{\bar q\to g}(Q^2,k_\perp^2)\\
&
\hspace{-1cm}+H_{e^+e^-\to qg}(Q^2,Q^2)\, J_{q\to q}(Q^2,k_\perp^2)\, J_{g\to g}(Q^2,k_\perp^2)+H_{e^+e^-\to g\bar q}(Q^2,Q^2)\, J_{g\to q}(Q^2,k_\perp^2)\, J_{\bar q\to g}(Q^2,k_\perp^2)\,.\nonumber
\end{align}
Again, exploiting charge conjugation, this simplifies to
\begin{align}
\sigma_{e^+e^-\to qg} &\supset H_{e^+e^-\to q\bar q}(Q^2,Q^2)\, J_{q\to q}(Q^2,k_\perp^2)\, J_{q\to g}(Q^2,k_\perp^2)\\
&
\hspace{1cm}+H_{e^+e^-\to qg}(Q^2,Q^2)\left[J_{q\to q}(Q^2,k_\perp^2)\, J_{g\to g}(Q^2,k_\perp^2)+ J_{g\to q}(Q^2,k_\perp^2)\, J_{q\to g}(Q^2,k_\perp^2)\right]\,.\nonumber
\end{align}

The final example we consider is the case in which a single quark transforms into any different flavor of quark, $\sum_{q'}\sigma_{e^+e^-\to qq'}$, such that the WTA flavor in the hard and jet functions differs, $q'\neq\bar q,q$.  We have
\begin{align}
\sum_{q'}\sigma_{e^+e^-\to qq'} &\supset \sum_{q'}H_{e^+e^-\to q\bar q}(Q^2,Q^2)\, J_{q\to q}(Q^2,k_\perp^2)\, J_{\bar q\to q'}(Q^2,k_\perp^2)\\
&
\hspace{1cm}+\sum_{q'}H_{e^+e^-\to qg}(Q^2,Q^2)\, J_{q\to q}(Q^2,k_\perp^2)\, J_{g\to q'}(Q^2,k_\perp^2)\nonumber\\
&\hspace{1cm}+\sum_{q'}H_{e^+e^-\to g\bar q}(Q^2,Q^2)\, J_{g\to q}(Q^2,k_\perp^2)\, J_{\bar q\to q'}(Q^2,k_\perp^2)\,.\nonumber
\end{align}
Exploiting flavor symmetry, this simplifies to
\begin{align}
\sum_{q'}\sigma_{e^+e^-\to qq'} &\supset 2(n_f-1)H_{e^+e^-\to q\bar q}(Q^2,Q^2)\, J_{q\to q}(Q^2,k_\perp^2)\, J_{q\to q'}(Q^2,k_\perp^2)\\
&
\hspace{1cm}+2(n_f-1)H_{e^+e^-\to qg}(Q^2,Q^2)\,J_{g\to q}(Q^2,k_\perp^2)\left[ J_{q\to q}(Q^2,k_\perp^2)+J_{q\to q'}(Q^2,k_\perp^2)\right]\,.\nonumber
\end{align}

We plot these WTA flavor cross section predictions as a function of the center-of-mass collision energy in $e^+e^-\to $ hadrons events in \Fig{fig:eexsec}.  For the LL prediction, we have multiplied the predicted flavor cross section by the total cross section $\sigma_\text{tot}$ so that the LL and LL$'$ predictions have the same normalization.  On these plots, we have included representative scale variation bands, corresponding to varying the natural scale $\mu_J$ of the jet functions by a factor of 2 around $k_\perp = 2$ GeV.  Going from LL to LL$'$ dramatically reduces the scale sensitivity, which can be understood from the fixed-order expansion of the jet function in \Eq{eq:gensol2loop}.  Inclusion of the one-loop constants $c^{(1)}$ explicitly eliminates the single-logarithmic terms proportional to $c^{(1)}$ that arise from variation of the scale $k_\perp^2$ at order-$\alpha_s^2$.  Further, there is a natural hierarchy in the WTA flavor fractions, $ \sigma_{e^+e^-\to qg}\sim \alpha_s\sigma_{e^+e^-\to q\bar q}$ and $ \sigma_{e^+e^-\to q q'}\sim \alpha_s\sigma_{e^+e^-\to qg}$, representative of the order in $\alpha_s$ at which the transition is first allowed.  Until nearly $Q = 20 $ GeV, the $\sigma_{e^+e^-\to q q'}$ cross section at LL$'$ accuracy is negative, which is a consequence of missing the complete low-scale matrix element $J_{q\to q'}\left(\alpha_s(\mu_J^2),\alpha_s(k_\perp^2)\right)$ that first contributes at ${\cal O}(\alpha_s^2)$.  At sufficiently high $Q$, the logarithms dominate, ensuring that the cross section is positive.

\section{Conclusions}\label{sec:concs}

A theoretically well-defined notion of jet flavor has important consequences for high precision calculations with exclusive heavy flavor production, for classification and identification of jets of different origins, and for constraining correlations between the final state and initial state for improved extractions of parton distribution functions.  We have extended the WTA flavor prescription of \InRef{Caletti:2022glq}, establishing a factorization \theorem~for jet flavor that is systematically improvable and can be incorporated with fixed-order results.  The evolution of flavor in the flow from the UV to the IR is governed by a modification to the DGLAP equations, and zero-bin subtraction is vital for ensuring that IR divergences are eliminated.  We have demonstrated consistency of the factorization \theorem~through two-loop order, and look forward to studies that push the precision boundary.

The calculations and factorization \theorem~presented here are just the beginning of the exploration of the utility of this WTA flavor definition.  In most practical use cases, jet flavor would be applied to production of heavy quarks, charm or bottom, and so quark masses will be important.  Here, all calculations have been presented for massless quarks, which would implicitly require that the IR scale is parametrically larger than the quark mass, $k_\perp^2 \gg m_q^2$.  However, relevant quark masses are on the scale of a few GeV, at the same order as the end of the parton shower, and so this massless flavor theory must match on to other theories that incorporate masses.  For high-energy jets, but assuming the scaling $k_\perp^2\sim m_q^2$, we would match onto a quasi-collinear theory, for which one-loop jet functions can be calculated from known results \cite{Catani:2000ef,Leibovich:2003jd}.  At even lower cutoffs, $k_\perp^2\ll m_q^2$, we would match onto a version of (boosted) heavy quark effective theory \cite{Eichten:1989zv,Georgi:1990um,Grinstein:1990mj,Mannel:1991mc,Fleming:2007qr}, and further may possibly need to incorporate heavy quark decays.

While the jet flavor is a purely perturbative notion, requiring partons on which to be defined, to be incorporated into real, experimental measurements would require an interface with hadronization.  Further, the exclusively collinear nature of the WTA flavor factorization \theorem~implies that this can be incorporated by another universal fragmentation function that would describe the transition of a parton of flavor $f$ at the scale $k_\perp^2$ into the hadron that lies along the measured WTA axis at a scale comparable to the QCD scale, $\Lambda_\text{QCD}^2$.  First studies into the distribution of hadrons about the WTA axis were presented in \InRef{Neill:2016vbi}.  Just as for the functions in the perturbative factorization \theorem, the scale dependence of this partonic-to-hadronic flavor transition would be governed by modified DGLAP equations, but there is a subtlety now with flavor.  Partonic flavor would have to be mapped to different hadron species, which muddies a short-distance interpretation of flavor, but would much more naturally connect with practical experimental definitions of heavy-flavor tagged jets.  In this way, these hadronic flavor fragmentation functions may have a close relationship with track functions \cite{Chang:2013rca,Chang:2013iba,Li:2021zcf,Jaarsma:2022kdd,Chen:2022pdu}, that describe the production of electrically-charged hadrons from partons.

A piece of the two-loop anomalous dimensions will involve next-to-leading order timelike splitting functions \cite{Curci:1980uw,Floratos:1981hs,Furmanski:1980cm,Ellis:1996mzs} separated into distinct flavor channels.  Some more recent studies have developed this further, especially for application in a next-to-next-to-leading logarithmic parton shower \cite{Ritzmann:2014mka,Hoche:2017iem,Hoche:2017hno}.  Splitting functions, however, measure the energy of a single particle in the splitting and are inclusive over all other particles, but the definition of the WTA axis requires re-clustering particles and measuring the hardest particle at each stage of the reconstructed branching history.  Thus, part of the two-loop WTA flavor anomalous dimensions will be the probability that a particle produced in the splitting will have energy fraction greater than 1/2, but clustering effects will be significant.  Nevertheless, a fruitful way forward to higher precision may be to first extract universal or clustering-independent anomalous dimensions, especially if clustering effects can be expressed as a purely subleading logarithmic effect.

Because the WTA flavor evolves due to collinear splittings, the flavor jet functions are universal.  Factorization of WTA flavor at hadron colliders like the LHC then only requires new calculations for the hard function that establishes the WTA jet flavor in the UV.  However, incorporating WTA flavor in a high-precision fixed-order calculation may be non-trivial for a couple of reasons.  First, the WTA flavor is not IRC safe, so standard subtraction methods, e.g., \Refs{Catani:1996vz,Frixione:1995ms,Nagy:2003qn,Gehrmann-DeRidder:2005btv,Catani:2007vq,Gaunt:2015pea,Boughezal:2015dva,Cacciari:2015jma,DelDuca:2016csb}, cannot immediately be employed out of the box.  Second, the flavor information of individual particles must be retained along with reclustering and WTA recombination throughout the calculation, which is more exclusive and complicates phase space boundaries.  Nevertheless, the factorization \theorem~requires that the divergences of the hard function are universal and exclusively collinear in nature, and so there may be a flavored subtraction scheme that can be used to isolate the finite contributions of the hard function.  We look forward to these and other applications where the flavor of a jet can be exploited and improve analyses to one's own taste.

\acknowledgments

A.L.~thanks Simone Caletti and Daniel Reichelt for comments and Ian Moult for discussions of flavor track functions.  A.L.~was supported in part by the UC Southern California Hub, with funding from the UC National Laboratories division of the University of California Office of the President. D.N.~was supported by the U.S. Department of Energy, Office of Science, Office of Nuclear Physics (NP) contract DE-AC52-06NA25396 at Los Alamos National Lab, and by the LDRD program of LANL.

\appendix

\section{Explicit Solutions for Quark Flavor with One-Loop Evolution}\label{app:quarkevowta}

We first present the jet functions for the gluon WTA flavor with one-loop running, which we can then take as input to the expressions for quark WTA flavor.  From \Sec{sec:allordssols}, and inputing one-loop values, we have
\begin{align}\label{eq:gWTAoneloopevo}
&J_{g\to g}\left(\alpha_s(\mu^2),\alpha_s(k_\perp^2)\right) =J_{g\to g}\left(\alpha_s(\mu_J^2),\alpha_s(k_\perp^2)\right)\, \left(
\frac{\alpha_s(\mu^2)}{\alpha_s(\mu_J^2)}
\right)^{\frac{2n_f\gamma_{g\to q}^{(1)}+\gamma_{q\to g}^{(1)}}{\beta_0}}\\
&\hspace{1cm}+\frac{\gamma_{q\to g}^{(1)}\,J_{g\to g}\left(\alpha_s(\mu_J^2),\alpha_s(k_\perp^2)\right)+2n_f\gamma_{g\to q}^{(1)}\,J_{q\to g}\left(\alpha_s(\mu_J^2),\alpha_s(k_\perp^2)\right)}{2n_f\gamma_{g\to q}^{(1)}+\gamma_{q\to g}^{(1)}}\left(1- \left(
\frac{\alpha(\mu^2)}{\alpha_s(\mu_J^2)}
\right)^{\frac{2n_f\gamma_{g\to q}^{(1)}+\gamma_{q\to g}^{(1)}}{\beta_0}}\right)\,,\nonumber\\
&J_{q\to g}\left(\alpha_s,\alpha_s(k_\perp^2)\right) =J_{q\to g}\left(\alpha_s(\mu_J^2),\alpha_s(k_\perp^2)\right)\, \left(
\frac{\alpha_s(\mu^2)}{\alpha_s(\mu_J^2)}
\right)^{\frac{2n_f\gamma_{g\to q}^{(1)}+\gamma_{q\to g}^{(1)}}{\beta_0}}\\
&\hspace{1cm}+\frac{\gamma_{q\to g}^{(1)}\,J_{g\to g}\left(\alpha_s(\mu_J^2),\alpha_s(k_\perp^2)\right)+2n_f\gamma_{g\to q}^{(1)}\,J_{q\to g}\left(\alpha_s(\mu_J^2),\alpha_s(k_\perp^2)\right)}{2n_f\gamma_{g\to q}^{(1)}+\gamma_{q\to g}^{(1)}}\left(1- \left(
\frac{\alpha(\mu^2)}{\alpha_s(\mu_J^2)}
\right)^{\frac{2n_f\gamma_{g\to q}^{(1)}+\gamma_{q\to g}^{(1)}}{\beta_0}}\right)\,.\nonumber
\end{align}
Connecting to the leading-logarithmic results of \InRef{Caletti:2022glq}, we would set $J_{g\to g}\left(\alpha_s(\mu_J^2),\alpha_s(k_\perp^2)\right) = 1$ and $J_{q\to g}\left(\alpha_s(\mu_J^2),\alpha_s(k_\perp^2)\right)=0$, which then agrees exactly with the results established there.  Note also that in this form, there are in general ${\cal O}(\alpha_s)$ corrections to the value of the fixed point as expressed exclusively in terms of ratios of anomalous dimensions, through the higher-order terms in $J_{g\to g}\left(\alpha_s(\mu_J^2),\alpha_s(k_\perp^2)\right)$, for example.  

Only considering one-loop running, the quark evolution equations take a much simpler form than the all-orders expression:
\begin{align}
\frac{\partial}{\partial \alpha_s}\left(\begin{array}{c}
J_{q\to q}\left(\alpha_s,\alpha_s(k_\perp^2)\right)\\
J_{\bar q\to q}\left(\alpha_s,\alpha_s(k_\perp^2)\right)\\
J_{q'\to q}\left(\alpha_s,\alpha_s(k_\perp^2)\right)\\
J_{g\to q}\left(\alpha_s,\alpha_s(k_\perp^2)\right)
\end{array}\right)&=-\frac{1}{\alpha_s\beta_0}\left(
\begin{array}{cccc}
\gamma_{q\to q}^{(1)} &0&0&\gamma_{q\to g}^{(1)}\\
0 &\gamma_{q\to q}^{(1)}&0 &\gamma_{q\to g}^{(1)}\\
0 &0&-\gamma_{q\to g}^{(1)} &\gamma_{q\to g}^{(1)}\\
\gamma_{g\to q}^{(1)} &\gamma_{g\to q}^{(1)}&2(n_f-1)\gamma_{g\to q}^{(1)} &-2n_f\gamma_{g\to q}^{(1)}
\end{array}
\right)\left(\begin{array}{c}
J_{q\to q}\left(\alpha_s,\alpha_s(k_\perp^2)\right)\\
J_{\bar q\to q}\left(\alpha_s,\alpha_s(k_\perp^2)\right)\\
J_{q'\to q}\left(\alpha_s,\alpha_s(k_\perp^2)\right)\\
J_{g\to q}\left(\alpha_s,\alpha_s(k_\perp^2)\right)
\end{array}\right)\,,
\end{align}
where $\beta_0$ is the one-loop $\beta$-function coefficient and we have set $\gamma_{q\to \bar q}^{(1)} = \gamma_{q\to q'}^{(1)} = 0$ in the anomalous dimension matrix.  Now, there are only two distinct non-zero eigenvalues of the matrix,
\begin{align}
\gamma_{q,\text{WTA}_1}^{(1)} &= \gamma_{g\to g}^{(1)}-\gamma_{q\to g}^{(1)}\,,\\
\gamma_{q,\text{WTA}_2}^{(1)} &= \gamma_{q\to q}^{(1)}\,.
\end{align}
Additionally, because the flavor structure is particularly simple through one-loop, we have the sum rule that
\begin{align}
1&=J_{g\to g}\left(\alpha_s(\mu^2),\alpha_s(k_\perp^2)\right)+\sum_q J_{g\to q}\left(\alpha_s(\mu^2),\alpha_s(k_\perp^2)\right)\\
&=J_{g\to g}\left(\alpha_s(\mu^2),\alpha_s(k_\perp^2)\right)+2n_f J_{g\to q}\left(\alpha_s(\mu^2),\alpha_s(k_\perp^2)\right)\,.\nonumber
\end{align}
With the explicit solution for the jet function $J_{g\to g}\left(\alpha_s(\mu^2),\alpha_s(k_\perp^2)\right)$ from \Eq{eq:gWTAoneloopevo}, the $g\to q$ jet function is then
\begin{align}
&J_{g\to q}\left(\alpha_s(\mu^2),\alpha_s(k_\perp^2)\right) \\
&=\frac{1}{2n_f}-\frac{J_{g\to g}\left(\alpha_s(\mu_J^2),\alpha_s(k_\perp^2)\right)}{2n_f}\, \left(
\frac{\alpha_s(\mu^2)}{\alpha_s(\mu_J^2)}
\right)^{\frac{2n_f\gamma_{g\to q}^{(1)}+\gamma_{q\to g}^{(1)}}{\beta_0}}\nonumber\\
&\hspace{1cm}-\frac{\gamma_{q\to g}^{(1)}\,J_{g\to g}\left(\alpha_s(\mu_J^2),\alpha_s(k_\perp^2)\right)+2n_f\gamma_{g\to q}^{(1)}\,J_{q\to g}\left(\alpha_s(\mu_J^2),\alpha_s(k_\perp^2)\right)}{2n_f(2n_f\gamma_{g\to q}^{(1)}+\gamma_{q\to g}^{(1)})}\left(1- \left(
\frac{\alpha(\mu^2)}{\alpha_s(\mu_J^2)}
\right)^{\frac{2n_f\gamma_{g\to q}^{(1)}+\gamma_{q\to g}^{(1)}}{\beta_0}}\right)\nonumber
\end{align}
This relationship between $J_{g\to q}\left(\alpha_s(\mu^2),\alpha_s(k_\perp^2)\right)$ and $J_{g\to g}\left(\alpha_s(\mu^2),\alpha_s(k_\perp^2)\right)$ only holds at one-loop order, because at higher loops, there are more IR flavors that need to be included in the sum rule.

With this jet function, it is then easy to determine the other jet functions.  For example, the quark flavor-conserving jet function satisfies the evolution equation
\begin{align}
\alpha_s\frac{\partial}{\partial \alpha_s}J_{q\to q}\left(\alpha_s,\alpha_s(k_\perp^2)\right) = -\frac{\gamma_{q\to q}^{(1)}}{\beta_0}\,J_{q\to q}\left(\alpha_s,\alpha_s(k_\perp^2)\right)-\frac{\gamma_{q\to g}^{(1)}}{\beta_0}\,J_{g\to q}\left(\alpha_s,\alpha_s(k_\perp^2)\right)\,.
\end{align}
The solution is
\begin{align}
&J_{q\to q}\left(\alpha_s(\mu^2),\alpha_s(k_\perp^2)\right) =\frac{\gamma_{q\to g}^{(1)}}{2n_f(2n_f\gamma_{g\to q}^{(1)}+\gamma_{q\to g}^{(1)})}\left(
1-\left(
\frac{\alpha_s(\mu^2)}{\alpha_s(\mu_J^2)}
\right)^{\frac{2n_f\gamma_{g\to q}^{(1)}+\gamma_{q\to g}^{(1)}}{\beta_0}}
\right)\\
&\hspace{2cm}-\frac{\gamma_{q\to g}^{(1)}J_{g\to g}\left(\alpha_s(\mu_J^2),\alpha_s(k_\perp^2)\right)-2n_f\gamma_{g\to q}^{(1)}J_{q\to q}\left(\alpha_s(\mu_J^2),\alpha_s(k_\perp^2)\right)}{2n_f(2n_f\gamma_{g\to q}^{(1)}+\gamma_{q\to g}^{(1)})}\nonumber\\
&\hspace{2cm}+\frac{2n_f-1}{2n_f}\,J_{q\to q}\left(\alpha_s(\mu_J^2),\alpha_s(k_\perp^2)\right)\left(
\frac{\alpha_s(\mu^2)}{\alpha_s(\mu_J^2)}
\right)^{\frac{\gamma_{q\to g}^{(1)}}{\beta_0}}\nonumber\\
&\hspace{2cm}
+\frac{\gamma_{q\to g}^{(1)}}{2n_f}\frac{J_{g\to g}\left(\alpha_s(\mu_J^2),\alpha_s(k_\perp^2)\right)+J_{q\to q}\left(\alpha_s(\mu_J^2),\alpha_s(k_\perp^2)\right)}{2n_f\gamma_{g\to q}^{(1)}+\gamma_{q\to g}^{(1)}}\left(
\frac{\alpha_s(\mu^2)}{\alpha_s(\mu_J^2)}
\right)^{\frac{2n_f \gamma_{g\to q}^{(1)}+\gamma_{q\to g}^{(1)}}{\beta_0}}
\nonumber\,.
\end{align}
For our numerical results, we also need the quark-flavor transition jet function, which satisfies effectively the same differential equation with one-loop evolution,
\begin{align}
\alpha_s\frac{\partial}{\partial \alpha_s}J_{q\to q'}\left(\alpha_s,\alpha_s(k_\perp^2)\right) = -\frac{\gamma_{q\to q}^{(1)}}{\beta_0}\,J_{q\to q'}\left(\alpha_s,\alpha_s(k_\perp^2)\right)-\frac{\gamma_{q\to g}^{(1)}}{\beta_0}\,J_{g\to q}\left(\alpha_s,\alpha_s(k_\perp^2)\right)\,,
\end{align}
because at one loop, $\gamma_{q\to g}^{(1)} = -\gamma_{q\to q}^{(1)}$.  Its solution is then
\begin{align}
&J_{q\to q}\left(\alpha_s(\mu^2),\alpha_s(k_\perp^2)\right) =\frac{\gamma_{q\to g}^{(1)}}{2n_f(2n_f\gamma_{g\to q}^{(1)}+\gamma_{q\to g}^{(1)})}\left(
1-\left(
\frac{\alpha_s(\mu^2)}{\alpha_s(\mu_J^2)}
\right)^{\frac{2n_f\gamma_{g\to q}^{(1)}+\gamma_{q\to g}^{(1)}}{\beta_0}}
\right)\\
&\hspace{2cm}-\frac{\gamma_{q\to g}^{(1)}J_{g\to g}\left(\alpha_s(\mu_J^2),\alpha_s(k_\perp^2)\right)-2n_f\gamma_{g\to q}^{(1)}J_{q\to q}\left(\alpha_s(\mu_J^2),\alpha_s(k_\perp^2)\right)}{2n_f(2n_f\gamma_{g\to q}^{(1)}+\gamma_{q\to g}^{(1)})}\nonumber\\
&\hspace{2cm}-\frac{J_{q\to q}\left(\alpha_s(\mu_J^2),\alpha_s(k_\perp^2)\right)}{2n_f}\left(
\frac{\alpha_s(\mu^2)}{\alpha_s(\mu_J^2)}
\right)^{\frac{\gamma_{q\to g}^{(1)}}{\beta_0}}\nonumber\\
&\hspace{2cm}
+\frac{\gamma_{q\to g}^{(1)}}{2n_f}\frac{J_{g\to g}\left(\alpha_s(\mu_J^2),\alpha_s(k_\perp^2)\right)+J_{q\to q}\left(\alpha_s(\mu_J^2),\alpha_s(k_\perp^2)\right)}{2n_f\gamma_{g\to q}^{(1)}+\gamma_{q\to g}^{(1)}}\left(
\frac{\alpha_s(\mu^2)}{\alpha_s(\mu_J^2)}
\right)^{\frac{2n_f \gamma_{g\to q}^{(1)}+\gamma_{q\to g}^{(1)}}{\beta_0}}
\nonumber\,.
\end{align}
because through ${\cal O}(\alpha_s)$, $J_{q\to q'}\left(\alpha_s(\mu_J^2),\alpha_s(k_\perp^2)\right) = 0$.

\section{Two-loop Soft Contributions}\label{app:two_loop_soft}

We analyze the matrix element at two-loops for the soft contribution to quark flavor production. Since the soft emissions do not move the WTA axis, we may identify this axis as one of the eikonal lines in the soft approximation, $n$, and we keep in the soft limit the definition of light-cone fraction as the energy fraction, though this will not affect our conclusions. The relevant matrix element can be lifted from Ref. \cite{Catani:1999ss}, and is given by:
\begin{align}
|M_{n\bar{n}\rightarrow q\bar{q}}|^2\propto \frac{2(n\cdot k_1+ n\cdot k_2)(\bar{n}\cdot k_1+ \bar{n}\cdot k_2)(k_1\cdot k_2)-(n\cdot k_1 \bar{n}\cdot k_2-n\cdot k_2 \bar{n}\cdot k_1)^2}{(n\cdot k_1+ n\cdot k_2)^2(\bar{n}\cdot k_1+ \bar{n}\cdot k_2)^2(k_1\cdot k_2)}\,.
\end{align}
The momenta $k_1, k_2$ are the momenta of the quarks. We have dropped factors of $n_f$, $\alpha_s$, and color as these are inessential to our analysis, and let  $E_J=1$, which can be restored via dimensional analysis. The relevant phase space region is given by:
\begin{align}
&\int d\Omega_{n\bar{n}\rightarrow q\bar{q}}\\
&\hspace{0cm}=\int [d^dk_1]_+[d^dk_2]_+\Theta\Big(\bar{n}\cdot k_2-\bar{n}\cdot k_1\Big)\Theta\Big(k_{1\perp}^2-k_T^2\Big)\Theta\Big(k_{2\perp}^2-k_{1\perp}^2\Big)\Theta\Big(\frac{k_1\cdot k_2}{\bar{n}\cdot k_1\bar{n}\cdot k_2}-\frac{k_{1\perp}^2}{(\bar{n}\cdot k_1)^2}\Big)\,,\nonumber\\
&[d^dk_i]_+=\frac{d^dk_i}{(2\pi)^{d-1}}\Theta(k_i^0)\delta(k_i^2)\,.
\end{align}
This is simply the phase-space constraint \Eq{eq:soft_phase_space}, written in the appropriate soft variables, with the additional assumption that $z_1<z_2$. Analyzing the case $z_2>z_1$ results in the same conclusions. Making use of the on-shell conditions of the final state particles, and rescaling as:
\begin{align}
z_{i}&=_{\rm def.}\bar{n}\cdot k_{i}\,, \\
n\cdot k_i&=\frac{k_{i \perp}^2}{z_i}\,, \\
z_{2}&\rightarrow z_1z_2\,,\\
k_{\perp 2}&\rightarrow k_{\perp 2}k_{\perp 1}\,.
\end{align}
We achieve:
\begin{align}
\int d\Omega_{n\bar{n}\rightarrow q\bar{q}}&=\Omega_{d-2}\int d\Omega_{12}\frac{dz_1 dk_{1_\perp} k_{1\perp}^{1-4\epsilon}}{(2\pi)^{3-2\epsilon}} \frac{dz_2 dk_{2_\perp} k_{2\perp}^{1-2\epsilon}}{(2\pi)^{3-2\epsilon}}\Theta\Big(z_2-1\Big)\Theta\Big(k_{1\perp}^2-k_T^2\Big)\Theta\Big(k_{2\perp}^2-1\Big)\nonumber\\ & \qquad\qquad\times\Theta\Big( k_{2\perp}^2-z_2^2-2z_2k_{2\perp}\cos\phi_{12}\Big)\,,
\end{align}
The angle $\phi_{12}$ is the relative angle of the momenta projected into the transverse plane, i.e., the angle between $\vec{k}_{1\perp}$ and $\vec{k}_{2\perp}$, and $d\Omega_{12}$ is the appropriate phase space measure. The integral over $z_1$ is now unconstrained by the phase space, and the matrix element takes the form:
\begin{align}
|M_{n\bar{n}\rightarrow q\bar{q}}|^2\propto \frac{z_2^2k_{2\perp}\big(k_{2\perp}^4+z_2^2-2(1+z_2)k_{2\perp}(z_2+k_{2\perp}^2)\cos\phi_{12}+k_{2\perp}^2(1+4z_2+z_2^2)\big)}{z_1k_{1\perp}(1+z_2^2)(z_2+k_{2\perp}^2)^2(k_{2\perp}^2+z_2^2-2z_2k_{2\perp}\cos\phi_{12})^2}\,.
\end{align}
The integral over $z_1$ is divergent, but upon appropriate regularization with a rapidity regulator \cite{Chiu:2012ir}, is scaleless. We can repeat the analysis without using the light-cone approximation to the energy fractions. This leads to a more complicated matrix element, but, the energy fraction integral over $z_1$, after the appropriate rescalings, is again unconstrained and now regulated in dimensional regularization, and we conclude it is scaleless. We can repeat this analysis for the phase space in \Eq{eq:soft_phase_space_II}, once more finding scaleless integrals.

\section{Summary of One-Loop Results}\label{app:oneloopsum}

In this appendix, we summarize the one-loop calculations presented in \Sec{sec:uvrenormexp} for both the hard function for jet production in $e^+e^-$ collisions, and the jet function that describes collinear flavor transitions.  In these expressions, the coupling is evaluated at the scale $\mu^2$, $\alpha_s(\mu^2)$, and satisfies the one-loop $\beta$-function evolution
\begin{align}
\mu^2\frac{\partial \alpha_s}{\partial \mu^2} = -\frac{\alpha_s^2}{2\pi}\,\beta_0\,,
\end{align}
where the one-loop coefficient is
\begin{align}
\beta_0 = \frac{11}{6}C_A - \frac{2}{3}T_Rn_f\,.
\end{align}
The solution to the $\beta$-function is
\begin{align}
\alpha_s(\mu^2) = \frac{\alpha_s(m_Z^2)}{1+\frac{\alpha_s(m_Z^2)}{2\pi}\,\beta_0\log\frac{\mu^2}{m_Z^2}}\,,
\end{align}
where $m_Z = 91.18$ GeV and we use $\alpha_s(m_Z^2) = 0.118$.

\subsection{Hard Function}

The general form of the hard function $H_{e^+e^-\to j_L j_R}(Q^2,\mu^2)$ through one-loop accuracy is
\begin{align}
H_{e^+e^-\to j_L j_R} (Q^2,\mu^2)= \delta_{qj_L}\delta_{\bar qj_R} + \frac{\alpha_s}{2\pi}\left(
-\gamma_{e^+e^-\to j_Lj_R}^{(1)}
\log\frac{\mu^2}{Q^2}+h_{e^+e^-\to j_Lj_R}^{(1)}\right)+{\cal O}(\alpha_s^2)\,.
\end{align}
The relevant anomalous dimensions through this order are
\begin{align}
&\gamma_{e^+e^- \to q\bar q}^{(1)} =C_F\left(
\frac{5}{4}-4\log 2
\right)\,, &\gamma_{e^+e^-\to qg}^{(1)}=\gamma_{e^+e^-\to g\bar q}^{(1)} = C_F\left(
-\frac{5}{8}+2\log 2
\right)\,.
\end{align}
The low-scale constants are
\begin{align}
h_{e^+e^-\to q\bar q}^{(1)} &= C_F\left(-\frac{73}{24}+\frac{7}{2}\log 2-\frac{5}{4}\log 3+4\log^2 2+2\log^2 3+4\,\text{Li}_2\left(
\frac{2}{3}
\right)
\right)\,,\\
h_{e^+e^-\to qg}^{(1)}=h_{e^+e^-\to g\bar q}^{(1)} &= C_F\left(
\frac{109}{48}-\frac{7}{4}\log 2+\frac{5}{8}\log 3-2\log^2 2-\log^2 3-2\,\text{Li}_2\left(
\frac{2}{3}
\right)
\right)\,.
\end{align}

\subsection{Jet Function}

The general form of the jet function $J_{j\to f}(\mu^2,k_\perp^2)$ through one-loop accuracy is
\begin{align}
J_{j\to f}(\mu^2,k_\perp^2) &=\delta_{jf}+\frac{\alpha_s}{2\pi}\left( \gamma^{(1)}_{j\to f}\, \log\frac{\mu^2}{k_\perp^2}+c^{(1)}_{j\to f}\right)+{\cal O}(\alpha_s^2)\,.
\end{align}
The relevant anomalous dimensions through this order are
\begin{align}
&\gamma_{q\to q}^{(1)} = C_F\left(
\frac{5}{8}-2\log 2
\right)\,, &\gamma_{q\to g}^{(1)} = C_F\left(
-\frac{5}{8}+2\log 2
\right)\,,\\
&\gamma_{g\to q}^{(1)} = \frac{1}{3}T_R\,, &\gamma_{g\to g}^{(1)} = -\frac{2}{3}n_f T_R\,.\nonumber
\end{align}
The low-scale constants are
\begin{align}
&c_{q\to q}^{(1)} = C_F\left(
2-\frac{7}{4}\log 2-2\log^2 2
\right)\,, &c_{q\to g}^{(1)} = C_F\left(
-2+\frac{7}{4}\log 2+2\log^2 2
\right)\,,\\
&c_{g\to q}^{(1)} = T_R\left(
\frac{17}{36}-\frac{2}{3}\log 2
\right)\,, &c_{g\to g}^{(1)} = n_f T_R\left(
-\frac{17}{18}+\frac{4}{3}\log 2
\right)\,.\nonumber
\end{align}

\bibliography{flavor_fact}

\end{document}